\documentclass[twocolumn,aps,pr,superscriptaddress,preprintnumbers,nofootinbib,10pt]{revtex4-2}

\usepackage{amsmath,amssymb}
\usepackage[dvipdf,dvips]{graphicx}
\usepackage{color}
\usepackage{hyperref}
\usepackage{url}
\usepackage{slashed}
\usepackage{subfigure}
\usepackage[usenames,dvipsnames]{xcolor}
\usepackage{amsmath}
\usepackage{amsfonts}
\usepackage{float} 
\usepackage{amssymb}
\usepackage{epsfig}
\usepackage{graphics}
\usepackage{euscript}
\usepackage{slashed}
\usepackage{epstopdf}
\usepackage[utf8]{inputenc}
\allowdisplaybreaks
\usepackage[normalem]{ulem}
\usepackage{pifont}
\usepackage{dsfont}
\usepackage{verbatim}
\usepackage{graphicx}
\usepackage{latexsym}
\usepackage{courier}
\usepackage{mathrsfs}
\usepackage{braket}
\usepackage{physics}
\usepackage{tikz-feynman}

\def \L{\mathcal{L}}

\def \G{G_{\rm N}}

\def \A{\mathcal{A}}
\def \tt{\textmd{TT}}
\def \l{\textmd{L}}
\def \t{\textmd{T}}
\def \Det{\textmd{Det}}
\def \sdiff{\textmd{SDiff}}
\def \diff{\textmd{Diff}}

\def \H{\mathcal{H}}
\newcommand{\hTT}{h^{\rm TT}}
\newcommand{\htr}{h^{\rm tr}}
\newcommand{\aT}{a^{\rm T}}
\newcommand{\aL}{a^{\rm L}}
\newcommand{\PT}{P^{\mathrm T}}
\newcommand{\PiTT}{\Pi^{\mathrm{TT}}}
\newcommand{\intp}{\int\!\frac{\dd^4p}{(2\pi)^4}}
\newcommand{\Order}{\mathcal O}
\newcommand{\uPGT}{\mathrm{uPGT}}
\newcommand{\PGT}{\mathrm{PGT}}

\newcommand{\Kp}{K_{+}}
\newcommand{\Km}{K_{-}}

\newcommand{\Tbar}{\bar{\mathcal T}}

\newbox{\ORCIDicon}
\sbox{\ORCIDicon}{\large
	\includegraphics[width=0.8em]{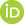}}

\hypersetup{
	colorlinks=true,
	citecolor=blue,
	citebordercolor=red,
	linktoc=all,
	linkcolor=blue,
	urlcolor=blue
}

\begin{document}
	
	\title{Towards Perturbative Unimodular Poincaré Gauge Theories with Propagating Torsion		}
	
	\author{Arthur F. Vieira\,\href{https://orcid.org/0000-0003-2897-2437}{\usebox{\ORCIDicon}}} \email{arthur.fvieira@gmail.com}
	\affiliation{UERJ $–$ Universidade do Estado do Rio de Janeiro,	Instituto de Física $–$ Departamento de Física Teórica $–$ Rua São Francisco Xavier 524, 20550-013, Maracanã, Rio de Janeiro, Brazil}
	
	\author{Antonio D. Pereira\,\href{https://orcid.org/0000-0002-6952-2961}{\usebox{\ORCIDicon}}} \email{adpjunior@id.uff.br}
	\affiliation{Instituto de F\'isica, Universidade Federal Fluminense, Campus da Praia Vermelha, Av. Litor\^anea s/n, 24210-346, Niter\'oi, RJ, Brazil}
	
	\author{Reinhard Alkofer\,\href{https://orcid.org/0000-0001-7433-3295}{\usebox{\ORCIDicon}}}\email{reinhard.alkofer@uni-graz.at}
	\affiliation{Institute of Physics, University of Graz, NAWI Graz, Universit\"atsplatz 5, 8010 Graz, Austria}

	\begin{abstract}
		We study the one-loop relation between an effective-field-theory extension of Poincaré gauge theories of gravity with propagating torsion and its unimodular counterpart. In the Einstein representation, we consider two representative quadratic torsion subsectors, namely totally	antisymmetric (axial) torsion and the vector component of hook-antisymmetric torsion. On maximally symmetric metric backgrounds with vanishing background torsion, the diffeomorphism-invariant and genuine unimodular theories yield identical one-loop determinants in both sectors, reproducing the same local logarithmic divergences. We further examine flat backgrounds with homogeneous axial or vector torsion, where metric-torsion mixing is present. In each case, the torsion-dependent one-loop effective action in unimodular gauge and in the genuine unimodular formulation is controlled by the same physical determinant. Thus, for the actions and backgrounds considered, the unimodular constraint does not modify local one-loop torsion dynamics. We also delineate the limitations of this equivalence and the extensions needed for generic torsionful curved backgrounds. 
		
	\end{abstract}
	\maketitle
	

	\section{Introduction} 
	\label{sec:Introduction} 
	
	More than a century after its formulation, General Relativity (GR) remains the most successful classical description of spacetime dynamics, with strong-field predictions confirmed by gravitational-wave observations and black-hole imaging \cite{Will:2014kxa,LIGOScientific:2016lio,EventHorizonTelescope:2020qrl}. Current observations are broadly consistent with GR, confirming its status as the benchmark theory of gravity.
	
	Nevertheless, both phenomenological and theoretical considerations motivate extensions of GR. Modified gravity has been explored in connection with dark matter, dark energy, and the Hubble tension \cite{Cooley:2022ufh,Arbey:2021gdg,Boveia:2022syt,Akerib:2022ort,Essig:2022dfa,Carney:2022gse,Baxter:2022dkm,Aramaki:2022zpw,Ando:2022kzd,Kahn:2022kae,Leane:2022bfm,Mitridate:2022tnv,Steigman:1985cco,SupernovaCosmologyProject:1998vns,SupernovaSearchTeam:1998fmf,Uzan:2006mf,Slosar:2019flp,Mortonson:2013zfa,Huterer:2017buf,Cortes:2026yde,RoyChoudhury:2026yyk}, although the latter may still reflect unresolved systematics. More fundamentally, perturbative GR faces well-known obstacles as a quantum field theory, motivating the search for consistent quantum extensions of gravity.
	
	Forthright power counting already signals the perturbative non-renormalizability of GR. In $d$ dimensions, Newton’s constant $\G$ has mass dimension $2-d$, and hence dimension $-2$ in four dimensions. Higher-loop corrections therefore require counterterms with progressively more derivatives, which cannot be absorbed into a finite set of couplings. Explicit calculations confirm this expectation: matter-coupled gravity develops non-renormalizable divergences at one loop, whereas pure GR without a cosmological constant is on-shell finite at one loop \cite{tHooft:1974toh,Christensen:1979iy}. With a cosmological constant, pure GR remains on-shell renormalizable at this order. At two loops, however, pure gravity generates a higher-curvature divergence that cannot be removed by field redefinitions or terms proportional to the classical equations of motion \cite{Goroff:1985th,Goroff:1985sz}. Thus, as a fundamental perturbative quantum field theory, GR requires an infinite tower of independent counterterms and loses predictive power as a fundamental theory at arbitrarily high energies.
	
	Yet this should not be misinterpreted as an incompatibility between GR and quantum field theory. In fact, the perturbative quantization of GR admits a consistent interpretation as an effective field theory (EFT) of quantum gravity. In the presence of an ultraviolet (UV) cutoff $\Lambda_{\rm UV}$, the infinite tower of higher-dimensional operators is systematically suppressed by inverse powers of $\Lambda_{\rm UV}$. For physical processes characterized by energy scales $E\ll \Lambda_{\rm UV}$, quantum-gravitational corrections scale parametrically as powers of $(E/M_{\rm Pl})^\#$, $\#>0$, where $M_{\rm Pl}=1/\sqrt{8 \pi \G}$ denotes the (reduced) Planck mass, identified with the natural UV cutoff of the effective description. This power-law suppression ensures that only a finite number of operators contribute appreciably at low energies, thereby restoring predictivity within the EFT framework, see, e.g., \cite{Donoghue:1994dn,Donoghue:2022eay}. 
	
	As the energy scale approaches the UV cutoff, however, this suppression mechanism ceases to operate and the EFT description breaks down. In this regime, GR must be substituted by a more fundamental theory capable of consistently capturing the quantum dynamics of spacetime at Planckian energies. These considerations naturally motivate the exploration of alternative formulations of gravity that preserve the empirical success of GR while potentially offering a different perspective on its UV structure. 
	
	Possibly such an extended concept of gravity is provided by going beyond the purely metric, Riemannian setting. Theories based on non-Riemannian geometries, such as Metric-Affine Gravity (MAG) \cite{Hehl:1994ue,Blagojevic:2012bc}, where the metric $g_{\mu\nu}$ and the affine connection ${A^\alpha}_{\mu\nu}$ are the fundamentally independent dynamical variables are here candidates, in particular, as such generalized geometric frameworks arise naturally in attempts to quantize gravity and to couple it consistently to matter with intrinsic spin \cite{Hehl:1976kj,Shapiro:2001rz}. The pure metric formulation is particularly unsuitable for coupling fermions, which are fundamental components of the Standard Model of particle physics, to gravitational interactions.
	
	Metric-compatible MAG constitutes a particularly relevant subclass of non-Riemannian theories. In this framework, the affine connection is constrained by the metric-compatibility condition, $D_\alpha g_{\mu\nu}=0$, such that the torsion $T^{\alpha}_{\phantom{\alpha}\mu\nu}$ constitutes the geometric degree of freedom independent of the metric, with $D$ denoting the gauge-covariant derivative associated with the affine connection.\footnote{Metric-affine models with vanishing torsion but nonzero nonmetricity provide another important class of non-Riemannian gravitational theories. Such metric-incompatible geometries have attracted renewed attention in recent years, in particular in relation to symmetric teleparallel formulations, propagating nonmetricity, and higher-spin extensions of gravity \cite{Latorre:2017uve,Conroy:2017yln,Iosifidis:2018zjj,Iosifidis:2018zwo,BeltranJimenez:2019esp,Jarv:2018bgs,Delhom:2019wir,Shimada:2018lnm,Aoki:2019rvi,Percacci:2020ddy,Heisenberg:2023lru,Mikura:2024mji,Percacci:2025oxw}.} A distinguished gauge-theoretic subclass is Poincaré gauge theories (PGTs), which realize such Riemann-Cartan geometry by gauging the Poincaré group $ISO(1,3)$. In the conventional gauge-theoretic formulation, the typical gravitational Lagrangian is constructed from the Hilbert-Palatini term together with a Yang-Mills-type term in the Poincaré field strengths, namely the Lorentz curvature ${F^\alpha}_{\beta\mu\nu}$ and the torsion ${T^\alpha}_{\beta\mu}$, which constitute the corresponding Cartan variables. Schematically, it reads \cite{Mikura:2023ruz}
	\begin{equation}
		\L_{\rm PGT}=M_{\rm Pl}^2 F+a T^2+c F^2.\label{eq:standard-PGT}
	\end{equation}	
	
	Recent analyses indicate that even comparatively simple metric-affine models with propagating torsion and nonmetricity remain perturbatively nonrenormalizable at one loop, despite the inclusion of operators containing up to four derivatives \cite{Melichev:2023lwj,Melichev:2024hih}. A controlled class of theories with propagating torsion, exhibiting asymptotic freedom and an absence of tachyonic instabilities, was investigated in~\cite{Melichev:2025hcg}. 

	Within this broader landscape of modified gravitational frameworks, an outstanding paradigm for modifying the gravitational sector while preserving the functional form of the action is Unimodular Gravity (UG) \cite{Anderson:1971pn,vanderBij:1981ym,Buchmuller:1988yn,Buchmuller:1988wx,Unruh:1988in,Henneaux:1989zc,Unruh:1989db,Alvarez:2008zw,Alvarez:2015pla,ellis2011trace} (see also the status report \cite{Carballo-Rubio:2022ofy} and the monograph \cite{Alvarez:2023juz}). In contrast to geometric extensions like PGT, UG introduces a restricted variational principle by constraining the determinant of the metric to be fixed to a prescribed scalar density, i.e., $g\equiv \det g_{\mu\nu}= \omega^2$ with some fixed function $\omega (x)$. As generic diffeomorphisms (\diff) do not preserve the metric determinant, this leads to a reduction in the symmetry group of the theory to the residual gauge symmetry associated with special (or volume-preserving) diffeomorphisms, denoted as \sdiff. This subgroup is generated by transverse vectors $\epsilon_\t^{\mu}$, which satisfy the condition $\nabla_\mu \epsilon_\t^\mu=0$, where the covariant derivative is defined with respect to the unimodular metric $g_{\mu\nu}.$
	
	At the classical level, standard GR and UG are known to be \textit{almost equivalent}, see, e.g., \cite{Percacci:2017fsy}. Locally, the two theories yield identical equations of motion, and their difference resides exclusively in the status of a single global degree of freedom associated with an overall scaling mode of the metric. In UG, this mode is eliminated by construction, and, as a consequence, the cosmological constant arises as an integration constant rather than as a free parameter in the action, thereby decoupling the cosmological constant from the vacuum energy density.
	
	This near-equivalence can be understood by noting that the unimodularity condition may be imposed locally as a restriction on the metric variations in GR. However, the equivalence is not exact. The unimodularity constraint implies the global identity\footnote{It is assumed that the manifold is compact or that an appropriate regularization is implemented.}%
	\begin{equation}
		V=	\int {\rm d}^4x\, \sqrt{|g(x)|}=\int {\rm d}^4x \,\omega(x)\,,
	\end{equation}%
	fixing the value of a physical observable and therefore cannot be interpreted as a mere gauge-fixing condition. As a result, GR possesses one additional global degree of freedom relative to UG. Classical UG is thus equivalent to GR formulated with fixed total volume.
	
	Beyond the classical theory, a longstanding debate exists in the literature regarding the persistence of this near-equivalence when quantum fluctuations are considered, see, e.g.,~\cite{deLeonArdon:2017qzg,Percacci:2017fsy,Fiol:2008vk,Saltas:2014cta,Padilla:2014yea,Smolin:2009ti,Smolin:2010iq,Alvarez:2015sba,Bufalo:2015wda,Upadhyay:2015fna,Eichhorn:2013xr,Eichhorn:2015bna,Benedetti:2015zsw,Herrero-Valea:2018ilg,DeBrito:2019gdd,deBrito:2020rwu,deBrito:2020xhy,Herrero-Valea:2020xaq,Kugo:2021bej,Kugo:2022iob,Kugo:2022dui}. In particular, a formal path-integral based proof hinting towards equivalence was presented in \cite{deBrito:2021pmw} for a general diffeomorphism-invariant action {$S_{\rm Diff}[g_{\mu\nu}]$}. There, the action is defined in terms of an unconstrained metric field, and the equivalence is obtained by gauge-fixing the longitudinal diffeomorphisms and by adopting a suitable definition of the functional measure. Under these conditions, the path integral of the fully diffeomorphism-invariant theory was shown to coincide with that of its unimodular counterpart. 
	
	Despite these circumstances, the literature on the equivalence between standard gravity and UG remains sparse,  especially beyond the purely Riemannian (metric) setting. In particular, systematic analyses in non-Riemannian geometries are lacking at both the classical and quantum levels. At the classical level, however, the extension is relatively straightforward: a fully dynamical affine connection, independent of the metric, can be consistently interpreted as an additional set of matter fields. Within this framework, the equivalence has been explicitly established for unimodular Einstein-Cartan gravity \cite{Bonder:2018mfz,Corral:2018hxi}.
	
	At the quantum level, the extension of the formal proof of \cite{deBrito:2021pmw} to the Riemann-Cartan setting seems structurally straightforward. The unimodularity condition fixes only the metric determinant, thereby reducing the gauge symmetry from \diff\ to \sdiff, while leaving the torsion tensor unconstrained. Furthermore, it still transforms covariantly under diffeomorphisms, $\delta_\epsilon {T^\alpha}_{\beta\gamma}=\L_\epsilon {T^\alpha}_{\beta\gamma}$. Under these conditions, the proof is expected to follow analogously to the Riemannian case, suggesting that the equivalence may persist even beyond perturbation theory. Nevertheless, an explicit perturbative verification might provide useful insights but is currently lacking.
	
	In this context it is important to note that the unimodularity constraint qualitatively changes the interaction structure of the corresponding quantum field theory. Since the volume element is fixed, its usual perturbative expansion around the background metric is absent. Equivalently, trace-type metric fluctuations do not generate additional interaction terms through $\sqrt{| g(x)|}$. As a result, vertices that would otherwise arise from the expansion of the determinant factor are not present, including part of the graviton-torsion interaction structure. This modifies the perturbative dynamics in a nontrivial way and may affect the quantum-level equivalence with standard gravity. Addressing this issue is the main goal of the present work. Clearly, the absence of sharp empirical discriminants between UG and standard GR motivates investigating such extensions within the broader framework of metric-affine geometry.
	
	Consistent with the preceding discussion, in this work, we specialize to metric-compatible MAGs while employing an EFT-extended PGT operator basis through canonical dimension $\Delta\leq 4$. Beyond the conventional terms displayed in Eq.~\eqref{eq:standard-PGT}, this basis contains derivative and interaction operators involving torsion which, although not part of the usual Yang-Mills-type definition of PGT, are generically expected in an effective description and under renormalization~\cite{Melichev:2023lwj}. Schematically, with tensor indices suppressed\footnote{The coupling conventions adopted throughout this work follow the standardization introduced in~\cite{Mikura:2023ruz}.},	 
	\begin{align} \label{eq:PGTLagrangian-Cartan}
		\mathcal{L}_{\rm PGT}\big|_{\rm C} &= \frac 1 2 M^2_{\rm Pl} (F-2\Lambda)  + a T^2 + c^{FF}F^2 + c^{FT}F DT \nonumber \\
		&\quad + c^{TT}(DT)^2 + c^{FTT}FTT \nonumber \\
		&\quad + c^{TTT}TT DT + c^{TTTT}TTTT\,.
	\end{align}
	In a stricter terminology, the additional operators beyond $F^2$ and $T^2$ place Eq.~\eqref{eq:PGTLagrangian-Cartan} within the general metric-compatible, or antisymmetric, MAG class. Our use of the designation PGT should therefore be understood as shorthand for this EFT extension of the conventional Poincaré-gauge operator basis. The effective dynamics of the theory are parameterized by the coupling constants $a$ and the  coefficients $c^i$. Employing the post-Riemannian decomposition of the Lorentz connection,
	\begin{equation}\label{eq:post-r}
		{A^\alpha}_{\mu\nu} = {\Gamma^\alpha}_{\mu\nu} + \frac{1}{2} \left( {T^\alpha}_{\mu\nu} - {T_{\mu\nu}}^{\alpha} + {T_\nu}^{\ \alpha}{}_{\mu} \right)\,,
	\end{equation}
	where ${\Gamma^\alpha}_{\mu\nu}$ denotes the Levi-Civita connection, the curvature field strength $F$ admits a decomposition into the Riemannian curvature $R$ and torsion-dependent contributions, schematically, one has $F \sim R + \nabla T + T^2$. Here, $\nabla$ represents the covariant derivative associated with the Levi-Civita connection. We refer the reader to \cite{Mikura:2023ruz} for technical details.
	
	Treating torsion as an independent field, the Lagrangian \eqref{eq:PGTLagrangian-Cartan}, formulated in Cartan variables, can be recast, at least, at the classical level, into an Einstein-like form,
	\begin{align}\label{eq:PGTLagrangian-Einstein}
		\L_{\rm PGT}\big|_{\rm E}&=\frac 1 2 M^2_{\rm Pl} (R-2\Lambda)+ a T^2 +b^{RR}R^2+b^{RT}R \nabla T\nonumber\\
		&+b^{TT}(\nabla T)^2+b^{RTT}RTT\nonumber\\
		&+b^{TTT}TT \nabla T+b^{TTTT}TTTT\,,
	\end{align} 
	making explicit the higher-derivative interactions induced by torsion and its coupling to curvature. The mapping between the coefficients $c^i$ and the secondary parameters $b^i$ is provided in detail in~\cite{Mikura:2023ruz}.
	
	In the Einstein representation, the quadratic torsion sector naturally defines a mass matrix for the irreducible torsion components. From an EFT viewpoint, two regimes are then possible. In  a conservative scenario, all torsion eigenmodes have masses of order $M_{\rm Pl}$, i.e., the effective UV cutoff of the gravitational EFT. They therefore decouple from the low-energy spectrum, and the torsionless condition $T^\alpha{}_{\mu\nu}=0$ emerges dynamically rather than being imposed kinematically. In this regime, PGT reduces effectively to the metric EFT of gravity at accessible scales.
	
	A more interesting possibility is that some torsion eigenmodes are parametrically lighter than $M_{\rm Pl}$. The derivative torsion operators may then support genuine propagating degrees of freedom within the EFT domain of validity. This regime is, however, strongly constrained by the particle content of the linearized theory. In particular, special relations among the quadratic couplings can isolate ghost- and tachyon-free propagating modes around Minkowski spacetime \cite{Mikura:2023ruz}. This motivates our choice of the axial and vector torsion sectors below as representative Lagrangians, while the status of the corresponding coupling relations beyond the classical approximation will be discussed separately.
	
	The primary goal of this work is the derivation of the one-loop effective actions for both Diff-invariant PGT and its unimodular extension (uPGT) in the presence of dynamical torsion, focusing on the logarithmic UV divergences. Metric-compatible MAGs with propagating torsion have been extensively investigated \cite{Sezgin:1981xs,Neville:1979rb,Hammond:1990aw,Saa:1996mt,Fabbri:2020drl,Baikov:1992uh,Tseytlin:1981nu,Hernaski:2009wp,Deveras:2010wq}, including perturbative analyses of their UV divergences \cite{Tseytlin:1981nu,Piva:2021nyj,Melichev:2025hcg}. Related Functional Renormalization Group studies include asymptotically safe Hilbert-Palatini gravity in an on-shell reduction scheme \cite{Gies:2022ikv} and nonperturbative beta functions for MAG with nonpropagating torsion and nonmetricity \cite{Pagani:2015ema,Reuter:2015rta}.
	
	While PGT is conventionally formulated in the Cartan representation, this approach tends to obscure the underlying quantum correspondence between the Diff-invariant and unimodular settings. Indeed, demonstrating the leading-order quantum equivalence is non-trivial when the theories are cast in their gauge-theoretic Cartan forms.\footnote{ Although this question lies beyond the scope of the present work, an investigation of the equivalence between the Cartan and Einstein representations at the quantum level would be interesting and is deferred to future work.} 
	To ensure analytical transparency and facilitate a direct comparison of the functional measures, we present the theory in the metric-affine Einstein representation. In this framework, the mapping between the two theories becomes manifest, allowing for an explicit evaluation of their respective one-loop structures. 
	
	This paper is organized as follows: In Sec. \ref{sec:PGT-1loop}, we formulate the one-loop analysis of PGTs in the Einstein representation, specify the class of PGTs used, decompose the torsion tensor into axial and hook-antisymmetric sectors, and derive the corresponding Hessians, gauge-fixing terms, ghost contributions, and one-loop partition function. In Sec. \ref{sec:OneLoopUPGTs}, we impose the unimodular constraint, identify the residual SDiff gauge structure, and show that the unimodular partition function has the same determinant structure as the Diff-invariant PGT partition function up to the field-independent gauge-group volume, leading to the equivalence of the one-loop effective actions. In Sec. \ref{sec:Conclusions}, we discuss the implications of this equivalence for PGTs with propagating torsion and outline possible extensions. Appendix \ref{app:trace_heatkernels} collects the expressions for the trace of heat-kernel coefficients and Appendix \ref{app:flat-four-hessians} the ones for the gauge-fixed flat-space Hessians with non-vanishing background torsion. Further technical details can be found in the appendices \ref{app:flat-reduced-operators} and \ref{app:flat-momentum-integrals}.


\section{One-loop analysis of PGTs}\label{sec:PGT-1loop}

\subsection{General framework and truncation}
Within the terminology adopted in this work, the EFT-extended PGT introduced in Eq.~\eqref{eq:PGTLagrangian-Cartan} is represented in Einstein variables by an action functional of the spacetime metric $g_{\mu\nu}$, the torsion tensor $T^\alpha{}_{\mu\nu}$, and their covariant derivatives,
\begin{equation}
	{S_{\rm PGT}=S_{\rm PGT}[g_{\mu\nu},T^\alpha{}_{\mu\nu}] }\, .
\end{equation}
The metric and torsion are treated as independent dynamical variables, reflecting the underlying Riemann-Cartan geometric structure. 
In the following, we will consider PGTs with full \diff~invariance and with SDiff invariance, i.e., the respective unimodular version.

The metric is dimensionless, while the torsion tensor carries mass dimension one, consistent with its interpretation as the antisymmetric part of an affine connection. These canonical dimensions play an important role in organizing the effective action according to operator dimensions.

The Einstein representation also makes transparent the relation with higher-derivative metric gravity: the complete dimension-four basis contains the usual curvature-squared sector together with additional torsional degrees of freedom and curvature-torsion interactions. In the one-loop analysis reported in this paper, however, we deliberately switch off the pure $R^2$ sector and the non-minimal curvature-torsion operators in order to isolate the effect of propagating torsion on the comparison between the Diff-invariant and unimodular formulations. Specifically, we consider that the action can be written in the form
\begin{equation}
	S_{\rm PGT}[g_{\mu\nu},T^\alpha{}_{\mu\nu}] = S_g[g_{\mu\nu}] + S_T[g_{\mu\nu},T^\alpha{}_{\mu\nu}] \, .
\end{equation}
As it will become apparent below, this separation substantially simplifies the evaluation of the one-loop effective action while retaining the essential dynamical features of the torsion sector. 

MAG generally possesses a rich spectrum of propagating degrees of freedom, and special choices of the quadratic couplings can eliminate ghost and tachyon instabilities at the level of the linearized classical theory \cite{Percacci:2020ddy,Mikura:2023ruz,Mikura:2024mji}; see also \cite{Sezgin:1979zf,Sezgin:1981xs,Lin:2018awc,Anselmi:2020opi,Piva:2021nyj}. An important qualification is that such relations among otherwise independent couplings need not be preserved by quantum corrections. Radiatively stable constructions therefore require the relevant restrictions to follow from an additional symmetry or another renormalization-group-stable mechanism rather than from an unprotected tuning of parameters. Examples of this strategy in MAG have been discussed in \cite{Marzo:2021iok,Marzo:2024pyn,Barker:2024goa,Barker:2025xzd,Barker:2025rzd,Barker:2025fgo}. The present work has a different objective: we use the axial and vector sectors identified in \cite{Mikura:2023ruz} to define representative propagating-torsion Lagrangians, but evaluate the one-loop comparison between the Diff-invariant and unimodular theories for generic nondegenerate values of their quadratic couplings. Consequently, our calculation should not be interpreted as establishing ghost or tachyon freedom of the quantum theory. In what follows, the theory is formulated in Euclidean signature.

Within this setting, the action of the PGT takes the standard gauge-fixed form 
\begin{equation}\label{eq:gauge.fixed.pgt.action}
	{S = S_{\rm PGT}}+ S_{\rm g.f.} + S_{\rm gh.} \, ,
\end{equation}
where $S_{\rm PGT}$ contains the dynamical gravitational terms, while $S_{\rm g.f.}$ and $S_{\rm gh.}$ denote the gauge-fixing and Faddeev-Popov ghost actions, respectively. Their explicit form will be specified later.

\subsection{Algebraic decomposition of the torsion tensor}
In the present work, we impose additional kinematical constraints on the torsion tensor. We consider two complementary sectors: (i) Totally antisymmetric torsion (axial sector) and (ii) hook-antisymmetric torsion, defined by the vanishing of the totally antisymmetric component. The torsion tensor is thus decomposed algebraically as
\begin{align}
	{T^\rho}_{\mu\nu}={H^\rho}_{\mu\nu}+{t^\rho}_{\mu\nu}\,,
\end{align}
where ${H^\rho}_{\mu\nu}$ is totally antisymmetric, and ${t^\rho}_{\mu\nu}$ is hook-antisymmetric-type and contains the vector and purely traceless tensor components. The first projection isolates axial torsion, while the second isolates purely vector/tensor torsion. In the following, we focus first on the totally antisymmetric sector, i.e., the axial torsion, and then within the hook-antisymmetric one on vector torsion. Treating both sub-sectors separately leads to significant technical simplifications while retaining a reasonable degree of generality.

\subsubsection{Totally antisymmetric torsion: axial sector}\label{sec:axial_case}

The restriction of torsion to its totally antisymmetric part isolates a well-defined subsector of the theory in which torsion is entirely encoded in a rank-three antisymmetric tensor. From the structural viewpoint, this restriction significantly simplifies the action by reducing the number of independent invariants that can be constructed purely from torsion. In particular, all terms involving the vector trace $T_\mu={T^\nu}_{\mu\nu}$ and the purely tensorial irreducible part identically vanish, leaving only contractions built from the antisymmetric component and its covariant derivatives.

Within this framework, the gravitational action functional restricted to the antisymmetric torsion sector is expressed as\footnote{For conciseness, we adopt the compact notation $\int_x \equiv \int {\rm d}^4x$ to denote the integration over the four-dimensional (Euclidean) spacetime manifold.} 
\begin{align}
	S_{\rm PGT}\Big|_H&=	\int_x\sqrt{g} \Bigg(
	\frac{M_{\rm Pl}^2}{2}\big(2\Lambda -R(g)\big)+m_1^2 H^{\alpha\mu\nu}H_{\alpha\mu\nu}\nonumber\\
	&+b_1\nabla^\alpha H^{\mu\nu\lambda}\nabla_\alpha H_{\mu\nu\lambda}+b_4\nabla_\mu H^{\mu\alpha\beta}\nabla_\nu {H^\nu}_{\alpha\beta}\Bigg).
\end{align}
The first two terms constitute the Einstein-Hilbert action, with $M_{\rm Pl}=1/\sqrt{8 \pi \G}$ denoting the reduced Planck mass and $\Lambda$ the cosmological constant. The remaining terms describe the axial torsion sector: $m_1$ is a mass-like parameter, while $b_1$ and $b_4$ control the independent two-derivative kinetic structures for the totally antisymmetric torsion field\footnote{We remark that actions of this type were already considered by Neville \cite{neville1978gravity}. No claim of novelty for this classical torsion model is intended here.}.  Throughout, we follow the coupling-label conventions of \cite{Mikura:2023ruz}.

In four dimensions, a totally antisymmetric rank-three tensor is Hodge-dual to a vector field. Accordingly, the torsion tensor can be represented in terms of an axial vector field $\A_\mu$ defined by
\begin{equation}
	\A_\mu=\dfrac{{\epsilon_{\mu\lambda}}^{\alpha\beta}}{\sqrt{g}}{T^\lambda}_{\alpha\beta}=\dfrac{{\epsilon_{\mu\lambda}}^{\alpha\beta}}{\sqrt{g}}{H^\lambda}_{\alpha\beta}.
\end{equation}
Expressed in terms of the axial vector field, the axial PGT action takes the form  
\begin{align}\label{eq:new-classical-action}
	S\Big|_\A&=\int_x \sqrt{g}\,\Biggl(\frac{M^2_{\rm Pl}}{2} \big(2\Lambda-R(g)\big)+\frac{1}{6}m_1^2 \A_\mu \A^\mu\nonumber\\
	&+\frac{1}{18}(3b_1+b_4)\nabla_\beta \A_\alpha \nabla^\beta \A^\alpha-\frac{1}{18}b_4(\nabla_{\alpha}\A_{\beta}\nabla^{\beta}\A^{\alpha})\Biggr)\nonumber\\
	&+S_{\rm g.f.} +S_{\rm gh.}\,.
\end{align}	
This form makes explicit that the totally antisymmetric torsion sector is dynamically equivalent to a massive axial vector field minimally coupled to curvature. 


To evaluate the one-loop effective action, we perform a background-field expansion of the action to second order in the fluctuation fields leading to an expression of the corresponding Hessian. The dynamical metric $g_{\mu\nu}$ is decomposed into a fixed background $\bar{g}_{\mu\nu}$ and fluctuations $h_{\mu\nu}$ through a general functional relation $g_{\mu\nu}=f_1(\bar{g},h)_{\mu\nu}$. An analogous split is applied to the axial torsion vector, $\A_\mu=f_2(\bar{\A},a)_\mu$, where $a_\mu$ denotes the corresponding fluctuation. Specifically, for the metric sector, we adopt the non-linear exponential parametrization see, e.g., \cite{Percacci:2017fkn}:
\begin{align}\label{eq:exp-param}
	g_{\mu\nu}&=\bar{g}_{\mu\alpha}{[\textmd{exp}(\kappa {h^.}_.)]^\alpha}_\nu\nonumber\\
	&=\bar{g}_{\mu\nu}+\kappa h_{\mu\nu}+\sum^\infty_{n=2}\frac{\kappa^n}{n!}h_{\mu\alpha_1}\cdots {h^{\alpha_{n-1}}}_\nu\,,
\end{align}
with $\kappa=2/M_{\rm Pl}$. In this convention, $\kappa$ is introduced to canonically normalize the fluctuations.
Conversely, the axial torsion sector is treated via a standard linear decomposition:
\begin{align}
	\A_\mu&=\bar{\A}_\mu+   a_\mu\,,\\ 
	a_\mu&=a^\t_\mu+\bar{\nabla}_\mu \frac{1}{\sqrt{\bar{\Delta}}}a^\l \label{hodgesplit}\,,
\end{align}
where $a^\t_\mu$ is the transverse axial fluctuation obeying $\bar{\nabla}^{\mu} a_{\mu}^\t=0$ and $a^\l$ is the longitudinal piece. The covariant derivative $\bar{\nabla}$ is defined with respect to the background metric $\bar{g}_{\mu\nu}$. For simplicity, we choose a Levi-Civita covariantly constant axial vector, $\bar \nabla_\mu \bar{\A}_\nu=0.$ For future use, the decomposition \eqref{hodgesplit} has a unit Jacobian, a property that will be explored at the level of the path integral.

In the following, we assume that the Riemannian part of the background geometry is a four-dimensional maximally symmetric space. Accordingly, the Levi-Civita background curvature tensors satisfy	
\begin{subequations}
	\begin{align}
		\bar{R}_{\mu\nu}&=\frac{\bar{R}}{4}\bar{g}_{\mu\nu},\label{eq:maximally_symmetric_space1}\\ \bar{R}_{\mu\nu\alpha\beta}&=\frac{\bar{R}}{12}(\bar{g}_{\mu\alpha}\bar{g}_{\nu\beta}-\bar{g}_{\mu\beta}\bar{g}_{\nu\alpha})\,\label{eq:maximally_symmetric_space2},
	\end{align}
\end{subequations}	
with constant scalar curvature, $\bar{\nabla}_\mu\bar{R}=0$. Compatibility with the maximal symmetry of the background metric implies a vanishing background axial torsion $\bar \A_\mu = 0$, since a nonvanishing vector would select a preferred direction and therefore break isotropy. Moreover, on a nonflat maximally symmetric space, a Levi-Civita covariantly constant axial vector, $\bar \nabla_\mu \bar{\A}_\nu=0$, implies $\bar \A_\mu = 0$ due to the integrability condition
\begin{equation}
	[\bar\nabla_\mu,\bar\nabla_\nu]\bar \A_\lambda=\bar R_{\lambda\sigma\mu\nu}\bar \A^\sigma=0\,,
\end{equation}
whenever $\bar R \neq 0$. For this background, the quadratic terms on the metric and torsion fluctuations decouple, and the corresponding fluctuation operators can be reduced to minimal Laplace-type form by means of the York decomposition. On a general curved background with nonvanishing axial torsion, by contrast, the Hessian contains background-dependent mixing terms between metric and torsion fluctuations, such as $h_{\mu\nu}\bar{\A}^\alpha \bar{\nabla}^\nu \bar{\nabla}_\alpha a^{\t\mu}$ and  $h^{\rm tr} \bar{\A}^\alpha \bar{A}^\beta \bar{\nabla}_\beta \bar{\nabla}_\mu {h_\alpha}^\mu$, where $h^{\rm tr}=\bar{g}^{\mu\nu}h_{\mu\nu}$. Such contributions are not of standard Laplace type and therefore obstruct a direct application of conventional covariant heat-kernel techniques for minimal second-order operators. These nonminimal and off-diagonal contributions can in principle be treated using off-diagonal heat-kernel techniques \cite{Barvinsky:1985an}, adapted to matrix-valued nonminimal operators. However, carrying out this analysis requires a substantially more involved treatment of the coupled metric-torsion fluctuation system and lies beyond the scope of the present work. Related perturbative computations in MAG with such structures can be found in~\cite{Sauro:2025sbt,Melichev:2025hcg}.

For the evaluation of the one-loop partition function, we then use the York decomposition for $h_{\mu\nu}$:
\begin{align}
h_{\mu\nu}&=h_{\mu\nu}^{\tt}+2\bar{\nabla}_{(\mu}\xi_{\nu)}\nonumber\\
&+\left(\bar{\nabla}_\mu\bar{\nabla}_\nu-\frac{1}{4}\bar{g}_{\mu\nu}\bar{\nabla}^2\right)\sigma+\frac{1}{4}\bar{g}_{\mu\nu}h^{\rm tr}\,.
\end{align}
Here, $h_{\mu\nu}^{\tt}$ denotes the transverse-traceless tensor mode, satisfying $\bar{\nabla}^{\mu} h_{\mu\nu}^{\tt}=0$ and $\bar{g}^{\mu\nu}h_{\mu\nu}^{\tt}=0$; $\xi_{\mu}$ is a transverse vector obeying $\bar{\nabla}^\mu \xi_\mu=0$ and $\sigma$ parametrizes the off-trace scalar sector. With a view towards the path integral, the York decomposition induces a nontrivial Jacobian that must be taken into account.

The resulting Hessian reads
\begin{align}
S_{\rm PGT}^{(2)}\Big|_\A=\frac{1}{2}\int_x \sqrt{\bar{g}}\,&\bigg(h^{\tt}_{\mu\nu}\mathcal{H}^{\tt}h^{\tt,\mu\nu}+\xi_\mu \H^{\xi\xi} \xi^\mu\nonumber\\
&+\sigma \H^{\sigma\sigma}\sigma+\sigma \H^{\sigma h}h^{\rm tr}\nonumber\\
&+ h^{\rm tr} \H ^{h\sigma}\sigma+h^{\rm tr} \H^{hh} h^{\rm tr}\nonumber\\
&+a^\t_\mu \H^{a^\t a^\t}a^{\t,\mu}+ a^\l \H^{a^\l a^\l}a^\l\bigg)\,,
\end{align}
where
\begin{subequations}
\begin{align}
	\H^{\tt}&=\frac{1}{2}\left( \bar{\Delta}+\frac{\bar{R}}{6}\right)\equiv\frac{1}{2}\Delta_{h^{\t\t}}\, ,\label{eq:Hessian_standard_axial.htt}\\
	\H^{\xi\xi}&=0\,,\\
	\H^{\sigma\sigma}&=-\frac{3}{16}\bar{\Delta}^2\left(\bar{\Delta}-\frac{\bar{R}}{3}\right)\equiv -\frac{3}{16}\bar{\Delta}^2\Delta_S\,,\label{eq:Hessian_standard_axial.sigma}\\
	\H^{\sigma h}&=-\frac{3}{16}\bar{\Delta}\left(\bar{\Delta}-\frac{\bar{R}}{3}\right)\equiv -\frac{3}{16}\bar{\Delta}\Delta_S\,,\\
	\H^{hh}&=-\frac{3}{16}\left(\bar{\Delta}+\frac{\bar{R}}{3}-\frac{8}{3}\Lambda\right)\equiv -\frac{3}{16}\Delta_{h^{\rm tr}}\,,\\
	\H^{a^\t a^\t}&=\frac{x_1}{18}\left(\bar{\Delta}+ \frac{3m^2_1}{x_1}+\frac{b_4}{x_1}\frac{\bar{R}}{4}\right)\equiv \frac{x_1}{18}\Delta_{a^\t}\,,\label{eq:Hessian_standard_axial.at}\\
	\H^{a^\l a^\l}&=\frac{b_1}{6}\left(\bar{\Delta}+\frac{m^2_1}{b_1}-\frac{\bar{R}}{4} \right)\equiv \frac{b_1}{6}\Delta_{a^\l}\label{eq:Hessian_standard_axial.al}\,,
\end{align}
\end{subequations}
with $\H^{\sigma h} = \H^{h\sigma}$ and we define the effective coupling $x_1=3b_1+b_4$. Here and in the following we employ the shorthand notation $\bar{\Delta}=-\bar{\nabla}^2$ for the background Laplacian. It is useful to distinguish the generic nondegenerate sector considered here from the special single-mode theories identified in \cite{Mikura:2023ruz}. For totally antisymmetric torsion, the classically healthy $1^+$ sector is obtained by setting $b_1=0$, whereas the corresponding $0^-$ sector requires $x_1=3b_1+b_4=0$, together with the appropriate sign conditions on the remaining kinetic and mass parameters. By contrast, the determinant expressions derived below assume $b_1\neq0$ and $x_1\neq0$, so that both operators in Eqs.~\eqref{eq:Hessian_standard_axial.at} and \eqref{eq:Hessian_standard_axial.al} are nondegenerate. The single-mode theories therefore constitute degenerate limits requiring a separate treatment and are not imposed as parameter restrictions in the present one-loop calculation.

\subsubsection{Hook antisymmetric case: purely vector torsion}
\label{sec:vector_torsion}

We now turn to the complementary kinematical sector in which torsion is of hook{-antisymmetric} type. In this case, the totally antisymmetric component is set to zero and the torsion tensor satisfies
\begin{equation}
t^{\mu}{}_{[\nu\rho]} = t^{\mu}{}_{\nu\rho} \,,
\qquad
t_{[\mu\nu\rho]} \equiv 
\frac{1}{3}\left(
t_{\mu\nu\rho}
+ t_{\rho\mu\nu}
+ t_{\nu\rho\mu}
\right)=0\,.
\label{eq:hook_constraints}
\end{equation}
These algebraic constraints eliminate the axial-vector component of torsion and restrict the theory to the vector and purely traceless tensor components.

As shown in \cite{Harst:2014vca,Mikura:2023ruz}, the absence of the totally antisymmetric sector reduces the number of independent invariants that can appear in the Lagrangian. Therefore, the torsion sector in \eqref{eq:gauge.fixed.pgt.action} simplifies to a smaller set of independent quadratic structures. This reduction is purely algebraic and does not rely on any background assumptions. Inspired by \cite{Mikura:2023ruz}, the corresponding action is
\begin{align}
S_{\rm PGT}\bigg|_{t}&=\int_x \sqrt{g}\,\Big(
\frac{M^2_{\rm Pl}}{2}\big(2\Lambda -R(g)\big) + m_1^2\, t_{\mu\nu\alpha}\, t^{\mu\nu\alpha}\nonumber\\
&+ m_3^2\, {t_\mu}^{\mu\alpha}\, {t_\nu}^\nu{}_{\alpha}+ b_1\, \nabla^{\alpha} t^{\mu\nu\lambda}\, \nabla_{\alpha} t_{\mu\nu\lambda}\nonumber\\
& + b_3\, \nabla^{\alpha} t_{\mu}{}^{\mu\beta}\, \nabla_{\alpha} t_{\lambda}{}^{\lambda}{}^{\beta}
+ b_4\, \nabla_{\mu} t^{\mu\alpha\beta}\, \nabla_{\nu} t^{\nu}{}_{\alpha\beta}\nonumber\\
&+b_7\, \nabla_{\mu} t^{\mu\alpha\beta}\, \nabla_{\nu} t_{\alpha}{}^{\nu}{}_{\beta} + b_8\, \nabla_{\mu} t^{\alpha\mu\beta}\, \nabla_{\alpha} t^{\nu}{}_{\nu\beta}\biggr)\,.		
\end{align}	
In the present work, we further restrict attention to the purely vector subsector. The torsion tensor then reduces to 
\begin{equation}
t_{\mu\nu\rho}=\frac{1}{3}\,(v_\rho g_{\mu\nu}-v_\mu g_{\nu\rho})\,,
\label{eq:pure_vector_torsion}
\end{equation}
so that the only independent torsional degree of freedom is the vector field $v_\mu$. This leads to the simplified action
\begin{align}\label{eq:new-classical-action_vector}
S\Big|_v&=\int_x \sqrt{g}\,\Biggl(
\frac{M^2_{\rm Pl}}{2}\big(2\Lambda-R(g)\big)  -\frac{1}{3}b_8\nabla_{\mu}v^\mu \nabla_\nu v^\nu\nonumber\\
&+ \left(\frac{2m_1^2}{3}+m_3^2\right) v_\mu v^\mu-\frac{1}{9}(b_7+2b_4)\nabla_\mu v_\nu \nabla^\nu v^\mu\nonumber\\
&+\frac{1}{9}(6b_1+9b_3+2b_4+b_7+3b_8)\nabla_\nu v_\mu \nabla^\nu v^\mu\Biggr) 
\nonumber\\
&+S_{\rm g.f.} +S_{\rm gh.}\,.
\end{align}	
We follow a procedure analogous to that employed in the axial sector by expanding the action \eqref{eq:new-classical-action_vector} to quadratic order in the fluctuation fields. To this end, the vector torsion field is decomposed into a background configuration $\bar{v}_\mu$ and a fluctuation $\varphi_\mu$, which is further decomposed into transverse and longitudinal components, i.e.,
\begin{equation}
v_\mu=\bar{v}_\mu+ \varphi_\mu, \quad \varphi_\mu=\varphi^\t_\mu+\bar{\nabla}_\mu \frac{1}{\sqrt{\bar{\Delta}}}\varphi^\l.
\end{equation}
The field decomposition employed herein possesses a unit Jacobian, ensuring the invariance of the functional integration measure. Within this framework, the gravitational contributions to the Hessian operator exhibit a structural equivalence to those of the axial torsion sector. Specifically, the vector torsion components are obtained by substituting the axial counterparts with the following operators:
\begin{subequations}
\begin{align}
	\H^{\varphi^\t \varphi^\t} &= \frac{1}{9}y_4 \left( \bar{\Delta} + 3\frac{y_2}{y_4} + \frac{y_3}{y_4} \frac{\bar{R}}{4} \right)\equiv  \frac{1}{9}y_4\Delta_{\varphi^\t}, \label{eq:Hessian_standard_vectorl.vt} \\
	\H^{\varphi^\l \varphi^\l} &= \frac{y_1}{3} \left( \bar{\Delta} + \frac{y_2}{y1} - \frac{y_1+b_8}{y_1} \frac{\bar{R}}{4} \right)\equiv \frac{y_1}{3}\Delta_{\varphi^\l}. \label{eq:Hessian_standard_vector.vl}
\end{align}
\end{subequations}
Here the vector background torsion is also zero, $\bar v_\mu=0$, which is imposed by a Levi-Civita covariantly constant vector in a maximally symmetric curvature background, so that no additional background-dependent mixing terms arise in this sector. The effective coefficients $y_i$ are defined as linear combinations of the theory's primary coupling constants $b_i$ and ``mass'' parameters $m_i$: 
\begin{subequations}
\begin{align}
	y_1 &= 2b_1+3b_3\,,\label{eq:coefficients_vector1} \\
	y_2 &= 2m^2_1 + 3m^2_3\,,\label{eq:coefficients_vector2} \\
	y_3 &= 2b_4 + b_7\,,\label{eq:coefficients_vector3} \\
	y_4 &= 6b_1 + 9b_3+2b_4+b_7+ 3b_8\nonumber\\
	&=3y_1+y_3+3b_8\,.\label{eq:coefficients_vector4}
\end{align}
\end{subequations}

In the next section, we discuss the explicit form of the gauge-fixing action and the underlying Faddeev-Popov ghost sector.

\subsection{Gauge-fixing and ghost sectors}
Consider an infinitesimal diffeomorphism generated by a vector field $\epsilon^\mu$, which can be uniquely decomposed into transverse and longitudinal components according to\footnote{For the present purposes, it suffices to consider the background covariant derivative.}
\begin{equation}\label{eq:epsilondecomp}
\epsilon^\mu=\epsilon^{\t\mu}+\bar{\nabla}^\mu \frac{1}{\sqrt{\bar \Delta}}\phi, \quad \bar{\nabla}_\mu \epsilon^{\t\mu}=0.
\end{equation}
Under this decomposition, the induced variations of the York variables naturally separate into gauge-invariant and gauge-variant sectors, namely
\begin{align}\label{eq:transformations}
\delta_\epsilon h_{\mu\nu}^{\t\t}&=0,\qquad	\delta_\epsilon \xi_\mu=\epsilon^\t_\mu,\nonumber\\
\delta_\epsilon \sigma&=\frac{2}{\sqrt{\bar \Delta}} \phi, \quad \delta_\epsilon h^{\rm tr}=-2\sqrt{\bar{\Delta}}\,\phi\,.
\end{align}
It is convenient to reparameterize the scalar sector in terms of the gauge-invariant combination
\begin{equation}
s=\bar{\Delta}\sigma+h^{\rm tr}\,.
\end{equation}
In terms of this variable, the gravitational scalar sector of the Hessian, both in the axial and in the vector cases, becomes diagonal and takes the form
\begin{align}
S_{\rm PGT}^{(2)}\bigg|^{\rm grav}_{\rm scalar}&=\frac{1}{2}\int_x \sqrt{\bar{g}}\,\bigg[-\frac{3}{16}s\left(\bar{\Delta}-\frac{\bar{R}}{3}\right)s\nonumber\\
&+\frac{1}{8}h^{\rm tr}(4\Lambda-\bar{R})h^{\rm tr}\bigg]\,.
\end{align}
This decomposition isolates the gauge-invariant scalar mode $s$ from the trace fluctuation $h^{\rm tr}$, which remains gauge dependent.

Gauge redundancies can be fixed by adding a covariant gauge-fixing term of the form
\begin{equation}
S_{\rm g.f.}=\frac{1}{2 \alpha}\int_x \sqrt{\bar{g}}\, \bar{g}^{\mu\nu}\mathcal{F}_\mu \mathcal{F}_\nu\,,
\end{equation}
where the gauge-fixing condition is defined as
\begin{equation}
\mathcal{F}_\mu\equiv{F_\mu}^{\alpha\beta}h_{\alpha\beta}=\left(\delta^{(\alpha}_\mu \bar{\nabla}^{\beta)} -\frac{1+\beta}{4}\bar{g}^{{\alpha\beta}}\bar{\nabla}_\mu\right)h_{\alpha\beta}\,.
\end{equation}
The gauge-fixing parameters $\alpha$ and $\beta$ will be kept arbitrary throughout the calculation. Upon use of the York decomposition and introducing the scalar variable
\begin{equation}\label{eq:chi}
\chi=\sigma+\frac{\beta}{(3-\beta)\bar{\Delta}-\bar{R}}s =\frac{(3\bar{\Delta}-\bar{R})\sigma+\beta h^{\rm tr}}{(3-\beta)\bar{\Delta}-\bar{R}}\,,
\end{equation}
the gauge-fixing action reads
\begin{align}
S_{\rm g.f.}&=\frac{1}{2\alpha}\int_x \sqrt{\bar{g}}\,\bigg[\xi_\mu \left(\bar{\Delta}-\frac{\bar{R}}{4} \right)^2\xi^\mu\nonumber\\
&+\frac{(3-\beta)^2}{16}\chi \bar{\Delta}\left(\bar{\Delta}-\frac{\bar{R}}{3-\beta} \right)^2\chi\bigg]\,.
\end{align}
We note that under the transformation \eqref{eq:transformations} the variable $\chi$ transforms exactly	in the same way as $\sigma$, $\delta_{\epsilon} \chi=2\bar{\Delta}^{-1/2}\phi$, such that $\xi_\mu$ and $\chi$ can be regarded as gauge degrees of freedom, i.e., they are bookkeeping variables for coordinate redundancy, not propagating physical modes.

Finally, the gauge-fixing procedure induces the standard Faddeev-Popov ghost action\footnote{We restrict the Faddeev-Popov ghost action up to quadratic terms in the fluctuations, which is enough for the purposes of the present work.},
\begin{equation}
S_{\rm gh}=\int_x \sqrt{\bar{g}}\, \bar{C}^\mu {F_\mu}^{\alpha\beta}\bar{\nabla}_{(\alpha}C_{\beta)}\,.
\end{equation}
We decompose the Faddeev-Popov ghost fields into transverse and longitudinal components according to
\begin{equation}\label{eq:ghost_decomposition}
C_\mu=C_\mu^\t+\bar{\nabla}_\mu \frac{1}{\sqrt{\bar{\Delta}}}C^\l
\end{equation}
and analogously for the anti-ghost field $\bar{C}_\mu$. Here, $C_\mu^\t$ satisfies $\bar{\nabla}^\mu {C}_\mu^\t=0$. This change of variables has trivial Jacobian and therefore does not induce any additional contribution to the functional measure. In terms of these variables, the ghost action diagonalizes and splits into independent transverse and longitudinal contributions,
\begin{align}
S_{gh}&=-\int_x \sqrt{\bar{g}}\, \bigg[\bar{C}^{\t,\mu}\left(\bar{\Delta}-\frac{\bar{R}}{4}  \right)C_\mu^\t\nonumber\\
&+\frac{(3-\beta)}{2}\bar{C}^\l\left( \bar{\Delta}-\frac{\bar{R}}{3-\beta} \right)C^\l\bigg]\,.
\end{align}
We are now in a position to compute the one-loop partition function and the corresponding one-loop effective action for both parametrizations of the torsion tensor.

\subsection{One-loop partition function for PGTs}
At one-loop order, the partition function is obtained by performing the Gaussian functional integration over the fluctuations around the chosen backgrounds. After gauge fixing, the resulting path integral factorizes into contributions from the metric fluctuations, the axial (or vector) torsion sector, the Faddeev-Popov ghosts, and the Jacobian associated with the York decomposition. In the axial torsion case, this reads
\begin{equation}
Z^{\rm axial}_{\rm PGT}={\rm e}^{-S_{\rm PGT}[\bar{g}]}\int ({\rm d}\Phi)_{\rm Diff}\, J_1 e^{-S_{\rm grav}^{(2)}[\bar{g};\Phi]}\,,
\end{equation}
where we define the multiplet of fluctuation fields $\Phi:=( h^\tt,\, \xi,\, \chi,\,  s,\,  h,\, a^\t,\, a^\l,\, \bar{C}^\t,\, \bar{C},\,  \bar{C}^\l,\,  C^\l)^t$ and $S_{\rm PGT}[\bar{g}]$ denotes the classical action evaluated on the background metric, since the background axial torsion is zero. $S_{\rm grav}^{(2)}[\bar{g};\Phi]$ contains all fluctuating fields $\Phi$ up to second order. The subscript Diff placed in the measure is to emphasize that it is invariant under BRST transformations associated with diffeomorphisms. Consequently, the one-loop partition function is expressed as a product of functional determinants of the corresponding minimal second-order operators.

Integrating out the transverse-traceless tensor $h_{\mu\nu}^{\tt}$, the \diff-invariant scalar combination $s$, the transverse axial torsion mode $a_\mu^{\t}$ and the longitudinal mode $a^\l$, yield the contributions 
\begin{align}
&\Det_{(2)}^{-1/2}\Delta_{h^{\t\t}}\,\Det_{(0)}^{-1/2}\Delta_S\,\Det_{(1)}^{-1/2}\Delta_{a^\t}\,\Det_{(0)}^{-1/2}\Delta_{a^\l}\,,
\end{align}
where the subscripts indicate the spin of the corresponding fields. Similarly, functional integration over $(\xi_\mu,\chi)$ yields
\begin{equation}
\Det_{(1)}\left(\bar{\Delta}-\frac{\bar{R}}{4} \right)^{-1}\Det_{(0)}^{-1/2}\bar{\Delta}\,\Det_{(0)}\left(\bar{\Delta}-\frac{\bar{R}}{3-\beta} \right)^{-1}\,.
\end{equation}
The Faddeev-Popov ghost action contributes an additional factor. Integrating over the transverse and longitudinal ghost fields $(\bar{C}^\t,C^\t,\bar{C}^\l,C^\l)$ gives
\begin{equation}
\Det_{(1)}\left( \bar{\Delta}-\frac{\bar{R}}{4}\right)\Det_{(0)}\left(\bar{\Delta}-\frac{\bar{R}}{3-\beta} \right)\,.
\end{equation}
Finally, the {Jacobian associated with the York decomposition contributes as} 
\begin{equation}\label{eq.Jacobian}
\Det_{(1)}^{1/2}\left(\bar{\Delta}-\frac{\bar{R}}{4} \right)\Det_{(0)}^{1/2}\bar{\Delta}\,\Det_{(0)}^{1/2}\left(\bar{\Delta}-\frac{\bar{R}}{3} \right)\,.
\end{equation}
Collecting all contributions, the one-loop partition function of the axial sector of PGT is obtained as
\begin{align}\label{eq:partitionfunction-PGT}
Z^{\rm axial}_{\rm PGT}&=V_{\rm Diff} {\rm e}^{-S_{\rm PGT}[\bar{g}]}\nonumber\\
&\times\frac{\sqrt{ \Det_{(1)}\Delta_{C^\t}}}{\sqrt{ \Det_{(2)}\Delta_{h^{\t\t}}}\sqrt{\Det_{(1)}\Delta_{a^\t}}\sqrt{\Det_{(0)}\Delta_{a^\l}}}\,,
\end{align}
where $\Delta_{C^\t}=\bar{\Delta}-\frac{\bar{R}}{4}$ is the transverse ghost operator.

The gauge-independence of the final result is verified by the complete cancellation of the parameters $\alpha$ and $\beta$. The factor $V_{\rm Diff}$, which formally denotes the infinite volume of the diffeomorphism group, emerges as a consequence of the residual background-gauge symmetry preserved by the Faddeev-Popov procedure and can be absorbed into the overall normalization of the path integral. From the partition function, we extract the one-loop effective action through the relation $Z^{\rm axial}_{\rm PGT}=e^{-\Gamma^{\rm axial}_{\rm PGT}}$. An analogous derivation for the purely vector torsion sector follows a computationally analogous path and is omitted here for brevity.

\section{One-loop unimodular PGTs}\label{sec:OneLoopUPGTs}
We now turn to the unimodular version of Poincaré gauge theories (uPGTs), in both totally antisymmetric and hook antisymmetric torsion cases. We focus first on the axial torsion case. The starting point is the same PGT action introduced in Eq.~\eqref{eq:new-classical-action}, supplemented with the unimodularity constraint on the spacetime metric,
\begin{equation}
\sqrt{\det g_{\mu\nu}}=\omega\,,
\end{equation}
where $\omega(x)$ is a fixed, non-dynamical scalar density. Accordingly, the unimodular PGT action is defined as the restriction of the PGT action to the unimodular metric configuration space,
\begin{equation}
S_{\rm uPGT}[g_{\mu\nu},\mathcal{A};\omega]
\equiv
S_{\rm PGT}[g_{\mu\nu},\mathcal{A}]
\big|_{\sqrt{g}=\omega}\,,
\end{equation}
so that the metric determinant is no longer a dynamical variable.

Since the determinant of the metric is fixed, infinitesimal variations of the metric must satisfy $g^{\mu\nu}\delta g_{\mu\nu}=0$, which enforces tracelessness of admissible metric variations. Under an infinitesimal coordinate transformation generated by a vector field $\epsilon^\mu$, the metric varies as $\delta g_{\mu\nu}=g_{\nu\alpha}\nabla_\mu \epsilon^\alpha+g_{\mu\alpha}\nabla_\nu \epsilon^\alpha$, and the unimodularity condition then implies $\nabla_\mu \epsilon^\mu=0$. The residual gauge symmetry therefore consists of special {(or volume-preserving)} diffeomorphisms, defining the subgroup \sdiff, whose infinitesimal generators are transverse vector fields $\epsilon_\t^\mu$, satisfying
\begin{equation}
\bar{\nabla}_\mu\epsilon_\t^\mu=0\,,
\end{equation}
where the relation $\nabla_\mu\epsilon_\t^\mu=\bar{\nabla}_\mu\epsilon_\t^\mu$ is valid for unimodular metrics.	Accordingly, uPGT is invariant only under \sdiff.

A technically convenient and conceptually transparent way to implement the unimodularity constraint is to adopt the exponential parametrization of the metric, cf.~Eq.~\eqref{eq:exp-param}, and restrict the metric fluctuation field to be traceless, {i.e.}
\begin{equation}
\bar{g}^{\mu\nu} h_{\mu\nu}=0\,.
\end{equation}
This prescription defines the so-called \textit{minimal formulation} of UG \cite{deLeonArdon:2017qzg}. With this choice, the condition $\sqrt{g}=\omega$ is satisfied identically to all orders in the fluctuation expansion {provided that $\sqrt{\bar{g}} = \omega$}. Moreover, the residual $\sdiff$ symmetry is preserved manifestly, and no additional constraints need to be imposed at the level of the functional integral.

Under the unimodularity constraint, the quadratic expansion of the action closely parallels that of Sec.~\ref{sec:PGT-1loop}. The only structural modification is the absence of contributions involving the trace mode $h$, which is eliminated by construction. As a consequence, the Hessian operator simplifies and receives contributions exclusively from the transverse-traceless metric sector, the scalar mode $\sigma$, and the transverse and longitudinal components of the axial torsion. Its explicit form coincides with the result obtained in Sec.~\ref{sec:axial_case}, cf.~Eqs.~\eqref{eq:Hessian_standard_axial.htt}, \eqref{eq:Hessian_standard_axial.sigma},\eqref{eq:Hessian_standard_axial.at},\eqref{eq:Hessian_standard_axial.al}, with all terms associated with the trace mode $h$ identically absent. This reflects the fact that unimodularity acts as a kinematical restriction on the field space, rather than modifying the dynamical structure of the remaining propagating modes.

Instead of employing the standard Faddeev-Popov procedure to explicitly divide out the volume of \sdiff, we adopt the geometrically motivated formulation of the functional integral over equivalence classes of geometries introduced by Mazur and Mottola. This approach, originally developed in the context of GR \cite{Mazur:1989by,Bern:1990bh,Mottola:1995sj}, was subsequently adapted to UG~\cite{Percacci:2017fsy,Ohta:2018sze} and more recently extended to MAG~\cite{Sauro:2024ujx}. Its defining feature is that gauge redundancies are isolated directly at the level of field space through the York decomposition of the metric fluctuations, while the associated ghost contributions arise naturally as Jacobian factors induced by the change of variables. As we show below, this framework provides a particularly transparent setting for establishing the formal one-loop equivalence between Diff-invariant and uPGTs. 

Within this formulation, the one-loop path integral of the axial sector of uPGT is given by
\begin{equation}
Z^{\rm axial}_{\rm uPGT}={\rm e}^{-S_{\rm uPGT}[{\bar{g}_{\mu\nu};\omega}]}\int ({\rm d} \Phi)_{\rm SDiff} J_1 e^{-S_{\rm grav}^{(2)}[\bar{g}_{\mu\nu},\Phi;\omega]}\,,
\end{equation}
where ${\omega}=\sqrt{\bar{g}}$ is the fixed background volume density, $({\rm d}\Phi)_{\rm SDiff}=({\rm d} h^\tt {\rm d}\xi\, {\rm d}\sigma\,{\rm d} a^\t\,{\rm d} a^\l)$ and $J_1$ denotes the Jacobian associated with the York decomposition, cf. Eq. \eqref{eq.Jacobian}.

A key structural simplification arises from the transformation properties of the vector field $\xi_\mu$ under an infinitesimal transverse diffeomorphism  generated by $\epsilon^\t_\mu$ \cite{Percacci:2017fsy,Ohta:2018sze},
\begin{equation}
\delta_{\epsilon^\t} \xi_\mu=\epsilon^\t_\mu\,.
\end{equation}
This relation allows one to identify $\xi_\mu$ as a coordinate along the gauge orbit through the background metric, while $h^\tt_{\mu\nu}$ and $\sigma$ parametrize gauge-invariant directions. Consequently, the functional integration over $\xi_\mu$ may be traded for an integration over the {generator of $\sdiff$}, $ ({\rm d}\xi) \mapsto ({\rm d}\epsilon^\t_\mu),$ making the separation between physical and gauge degrees of freedom manifest at the level of the path integral.

Performing the Gaussian integration over the remaining fluctuation fields yields the following determinant contributions:
\begin{align}
&\Det_{(2)}^{-1/2}\Delta_{h^{\t\t}}\Det_{(0)}^{-1}\bar{\Delta}\,\Det_{(0)}^{-1/2}\Delta_S\nonumber\\
\times&\,\Det_{(1)}^{-1/2}\Delta_{a^\t} \,\Det_{(0)}^{-1/2}\Delta_{a^\l}\,,
\end{align}
together with the Jacobian factor arising from the York decomposition, cf. Eq.~\eqref{eq.Jacobian}. Collecting all contributions, the one-loop partition function of the axial sector of uPGT takes the form
\begin{align}\label{eq:ZuPGT}
Z^{\rm axial}_{\rm uPGT}&= {\rm e}^{-S_{\rm uPGT}[\bar g_{\mu\nu};\omega]}\int ({\rm d}\epsilon^\t)\,\Det_{(0)}^{-1/2}\bar{\Delta}\nonumber\\
&\times\,\frac{\sqrt{ \Det_{(1)}\Delta_{C^\t}}}{\sqrt{ \Det_{(2)}\Delta_{h^{\t\t}}}\sqrt{\Det_{(1)}\Delta_{a^\t}}\sqrt{\Det_{(0)}\Delta_{a^\l}}}\,.
\end{align}
Since the volume of the \sdiff~gauge group must be independent of the dynamical metric, its functional definition requires the Jacobian associated with the transverse decomposition of the diffeomorphism parameter. Following \cite{Percacci:2017fsy,deLeonArdon:2017qzg}, one formally writes
\begin{align}
V_{\rm SDiff} &=\int ({\rm d} \epsilon)\delta(\bar{g}^{\mu\nu}\bar \nabla_\mu \epsilon_\nu) \nonumber\\
&= \int ({\rm d} \epsilon^{\rm T} {\rm d} \phi)\delta (\bar{\Delta}^{1/2}\phi)\nonumber\\
&= \int ({\rm d} \epsilon^\t) \left( \Det_{(0)} \bar{\Delta} \right)^{-1/2},
\end{align}
where, from Eq.~\eqref{eq:epsilondecomp}, it follows that $({\rm d}\epsilon)=({\rm d}\epsilon^\t {\rm d} \phi)$. This ensures that the gauge-group volume is defined in a metric-independent way.

Upon factorization of this volume from the path integral, the residual functional determinant structure is found to be structurally identical to that of the Diff-invariant axial PGT formulation. Since the one-loop effective action is defined via the relation
\begin{equation}
\Gamma^{\rm axial}_{\rm uPGT} = -\log Z^{\rm axial}_{\rm uPGT},
\end{equation}
the factor $V_{\rm SDiff}$ contributes only as a field-independent additive constant to the vacuum energy, which does neither affect the dynamics nor the renornalization and can be absorbed into the normalization factor. Therefore, from our construction, 
\begin{equation}
{\Gamma^{\rm axial}_{\rm uPGT}[\bar g_{\mu\nu};\omega]=\Gamma^{\rm axial}_{\rm PGT}[\bar g_{\mu\nu}]\Big|_{\sqrt{\bar{g}}=\omega}}\,.
\end{equation}
The one-loop effective action for the axial sector is then
\begin{align}
\Gamma^{\rm axial}_{\rm PGT}[\bar{g}_{\mu\nu}]&=S_{\rm PGT}[\bar{g}_{\mu\nu}]+\frac{1}{2}\log\Det_{(2)}\Delta_{h^{\t\t}}\nonumber\\
&+\frac{1}{2}\log\Det_{(1)}\Delta_{a^\t}+\frac{1}{2}\log \Det_{(0)}\Delta_{a^\l}\nonumber\\
&-\frac{1}{2}\log  \Det_{(1)}\Delta_{C^\t}.
\end{align}
Considering the UV structure of the theory, the logarithmically divergent part of the one-loop effective action is particularly interesting, since it is independent of the details of the UV regularization scheme \cite{Salam:1951sj}. It can be systematically evaluated through the application of standard covariant heat kernel expansion techniques \cite{Percacci:2017fkn,Buchbinder:2021wzv}, yielding
\begin{align}
\Gamma^{\rm axial}_{\rm PGT}\Big|_{\rm log}&=-\frac{1}{2}\frac{1}{16\pi^2}\log\frac{\Lambda^2_{\rm UV}}{\mu^2}\int_x \sqrt{\bar{g}}\big[\tr b_4(\Delta_{h^{\tt}})\nonumber\\
&+\tr b_4(\Delta_{a^\t})+\tr b_4(\Delta_{a^\l})-\tr b_4(\Delta_{C^\t})  \big],
\end{align}	
where $\Lambda_{\rm UV}$ is the UV cutoff and $\mu < \Lambda_{\rm UV}$ is a finite reference mass scale. For our choice of background configuration in four dimensions, the explicit form of the trace of the heat-kernel coefficients used in this work can be found in App. \ref{app:trace_heatkernels}, leading to
\begin{align}
\Gamma^{\rm axial}_{\rm PGT}\Big|_{\rm log}&=-\frac{1}{32\pi^2}\log\frac{\Lambda^2_{\rm UV}}{\mu^2}\int_x \sqrt{\bar{g}}\bigg[A_1
+B_1\bar R+C_1\bar R^2 \bigg],
\end{align}
where the coefficients $A_1, B_1, C_1$ are
\begin{subequations}
\begin{align}
	A_1&=\frac{m_1^4}{2b_1^2}+\frac{27m_1^4}{2x_1^2},\label{eq.A1}\\
	B_1&=\frac{9b_4m_1^2}{4x_1^2} - \frac{5m_1^2}{12b_1} -	\frac{3m_1^2}{4x_1},\\
	C_1&=\frac{49}{360}+\frac{3b_4^2}{32x_1^2}-\frac{b_4}{16x_1},
\end{align}
\end{subequations}
with $x_1=3b_1+b_4.$ By parity of reasoning, an analogous derivation for the purely vector torsion sector can be achieved. In that case, the logarithmically divergent contribution takes the form
\begin{align}
\Gamma^{\rm vec}_{\rm PGT}\Big|_{\rm log}&=-\frac{1}{32\pi^2}\log\frac{\Lambda^2_{\rm UV}}{\mu^2}\int_x \sqrt{\bar{g}}\bigg[
A_2+ B_2 \bar R +	C_2 \bar R^2\bigg],
\end{align}
with
\begin{subequations}
\begin{align}
	A_2&=\frac{27y_2^2}{2\left(3b_8+3y_1+y_3\right)^2}+\frac{y_2^2}{2y_1^2},\label{eq:A2}\\
	B_2&=-\frac{b_8y_2}{4y_1^2}-\frac{5y_2}{12y_1}+\frac{9y_2y_3}{4\left(3b_8+3y_1+y_3\right)^2}\nonumber\\
	&-\frac{3y_2}{4\left(3b_8+3y_1+y_3\right)},\\
	C_2&=\frac{49}{360}+\frac{b_8^2}{32y_1^2}+\frac{5b_8}{48y_1}+\frac{3y_3^2}{32\left(3b_8+3y_1+y_3\right)^2}\nonumber\\
	&-\frac{y_3}{16\left(3b_8+3y_1+y_3\right)},
\end{align}
\end{subequations}
where the parameters $y_i$ are given by Eqs. \eqref{eq:coefficients_vector1}, \eqref{eq:coefficients_vector2}, \eqref{eq:coefficients_vector3} and \eqref{eq:coefficients_vector4}. 

\section{Flat-space consistency checks with homogeneous background torsion}
\label{sec:flat-space-consistency}

The calculation on maximally symmetric metric backgrounds is complemented here by a local flat-space check with a homogeneous background torsion. We take
\begin{equation}
\bar g_{\mu\nu}=\delta_{\mu\nu},\qquad
\Delta=-\partial^2,
\qquad
\partial_\rho \Tbar_\mu=0,
\label{eq:flat-common-background-main}
\end{equation}
where $\Tbar_\mu$ denotes $\bar\A_\mu$ in the axial sector and $\bar v_\mu$ in the vector sector.

This setting is useful because all momentum-space contractions can be performed explicitly and the one-loop determinant can be compared in two implementations of unimodularity. The first is the ${\rm Diff}$-invariant theory evaluated in the unimodular gauge. The second is the genuinely unimodular theory, whose gauge group is ${\rm SDiff}$.

We remark that the following calculation is off-shell in the background torsion. A nonzero constant $\Tbar_\mu$ is admissible in flat Euclidean space, but it selects a preferred direction and therefore breaks the manifest $O(4)$ symmetry of the chosen background configuration with the corresponding consequences for otherwise conserved quantities. In the following, we assume $\Tbar^2\not =0$, however, without breaking isotropy or homogeneity. The expressions below should consequently be interpreted as the local expansion of the one-loop effective action in $\Tbar^2$, not as an assertion that the constant background torsion solves the classical field equations.

\subsection{Gauge reduction and common fluctuation space}
\label{subsec:flat-gauge-reduction}

In the ${\rm Diff}$-invariant formulation, the terminology \emph{unimodular gauge}, also called the \emph{unimodular physical gauge} \cite{Ohta:2015fcu,Percacci:2015wwa,Alkofer:2018fxj}, denotes a gauge choice inside the full ${\rm Diff}$-invariant configuration space. It should not be confused with a theory whose configuration space is unimodular from the outset. For the exponential split, $g_{\mu\nu}=\bar g_{\mu\alpha}\bigl[e^{\kappa h}\bigr]^\alpha{}_\nu ,$ the limit $\beta\to-\infty$ of the usual (flat-space) two-parameter gauge condition
\begin{equation}
F_\mu[h;\bar g]
=
\partial^\nu h_{\mu\nu}
-
\frac{1+\beta}{4}\partial_\mu h^{\rm tr}
\label{eq:flat-diff-gauge-condition}
\end{equation}
enforces $\partial_\mu h^{\mathrm{tr}}=0$. Hence, the trace is reduced to a spacetime constant, $h^{\mathrm{tr}}(x)=h_0$, and the determinant of the full metric is fixed locally up to an overall volume mode. In the present local one-loop calculation, the constant mode is excluded from the determinants and the total volume is held fixed. The Landau gauge $\alpha \rightarrow0$ ensures $h_0=0$ strongly in the nonzero-mode sector\footnote{At the level of the path integral, this is equivalent to removing the trace mode $h^{\rm tr}$ from the quadratic action at the cost of generating the associated ghost contribution $\sqrt{\Det(\Delta)}$.}, such that $ \Delta\sigma\rightarrow s$. 

Nonperturbative evidence for the equivalence of these two implementations in the purely metric setting was obtained in~\cite{deBrito:2020rwu} at the level of connected $n$-point correlation functions within the adopted assumptions. However, their result does not by itself establish the statement in the
presence of torsion. The calculation below supplies an explicit one-loop test of the latter point for a constant background torsion.

In the genuinely unimodular formulation, instead, the exponential parametrization and the unimodular constraint imply $h^{\rm tr}=0$ identically, and the residual gauge group is \sdiff.  The transverse York vector $\xi_\mu$ is then a coordinate along the \sdiff~gauge orbit. Consequently, $\xi_\mu$ parametrizes a pure \sdiff~gauge direction rather than an independent physical fluctuation. In the Landau-gauge limit, this gauge direction is fixed sharply and, at the quadratic level, $\xi_\mu$ decouples from the physical fluctuation modes.  Its Gaussian determinant cancels the transverse ghost determinant, while the residual York and group-volume factors are independent of $\Tbar_\mu$ (see App. \ref{app:flat-four-hessians} and \ref{app:flat-reduced-operators} for details).  Consequently, the physical part of the torsion-background-dependent part of the one-loop effective action is governed in both formulations by the same determinant,
\begin{align}
\Gamma^{(1)}_{\Tbar\text{-dep.}}
& =\frac12\Tr_{\Phi_{\rm phys}^{(\mathcal T)}}
\log \mathcal O_{\rm phys}(\Tbar),    
\label{eq:flat-common-reduced-determinant-main}
\end{align}
for the physical multiplet $\Phi_{\rm phys}^{(\mathcal T)}=\left(\hTT_{\mu\nu},s,\psi^{\rm T}_\mu,\psi^{\rm L}\right),$ with $\psi_\mu=a_\mu$ for axial torsion and $\psi_\mu=\varphi_\mu$ for vector torsion.

The reduced operator is expanded as
\begin{equation}
\mathcal O_{\rm phys}(\Tbar)
=\mathcal O_0+\mathcal N_1+\mathcal N_2+
\Order(\Tbar^3),
\qquad
\mathcal N_n=\Order(\Tbar^n),
\label{eq:flat-common-operator-expansion-main}
\end{equation}
so that
\begin{align}
\frac12\Tr\log\mathcal O_{\rm phys}
={}&\frac12\Tr\log\mathcal O_0
+\frac12\Tr(\mathcal G_0\mathcal N_2)
\nonumber\\
&-\frac14\Tr(\mathcal G_0\mathcal N_1\mathcal G_0\mathcal N_1)
+\Order(\Tbar^3),
\label{eq:flat-common-tracelog-main}
\end{align}
where $\mathcal G_0=\mathcal O_0^{-1}$.  The term linear in $\Tbar_\mu$ vanishes because $\mathcal G_0$ is block diagonal whereas $\mathcal N_1$ is off diagonal in field space.  The unreduced gauge-fixed Hessians, the sector-dependent reduced operators, and the momentum-space expansions are collected in Apps.~\ref{app:flat-four-hessians}-\ref{app:flat-momentum-integrals}.

For the axial sector, $\Phi_{\rm phys}^{\rm axial}=(\hTT_{\mu\nu},s,\aT_\mu,\aL)$, and the common logarithmically divergent contribution up to $\Order(\bar\A^2)$ is
\begin{equation}
\begin{aligned}
	\Gamma^{(1)}_{\log,\rm axial}
	={}&-
	\frac{V_4m_1^4}{64\pi^2}
	\log\frac{\Lambda_{\rm UV}^2}{\mu^2}
	\Bigg\{
	\frac{27}{(3b_1+b_4)^2}
	+\frac{1}{b_1^2}
	\\
	&\quad
	-\frac{\kappa^2\bar\A^2}{16}
	\left[
	\frac1{b_1}
	-\frac{9(3b_1+2b_4)^2}{(3b_1+b_4)^3}
	\right]
	\Bigg\}.
\end{aligned}
\label{eq:flat-final-axial-main}
\end{equation}
The first line is the curvature-independent part of the curved-space result, \eqref{eq.A1}, while the second line is the local off-shell contribution generated by the homogeneous axial background.

For the vector sector, $\Phi_{\rm phys}^{\rm vec}=(\hTT_{\mu\nu},s,\varphi^\t_\mu,\varphi^\l)$.  In terms of
\begin{align}
y_1&=2b_1+3b_3,
&y_2&=2m_1^2+3m_3^2,
\nonumber\\
y_3&=2b_4+b_7,
&y_4&=3y_1+y_3+3b_8,
\label{eq:vector-y-main}
\end{align}
we obtain
\begin{align}\label{eq:flat-final-vector-main}
\Gamma^{(1)}_{\log,\rm vec}
&=-
\frac{V_4y_2^2}{64\pi^2}
\log\frac{\Lambda_{\rm UV}^2}{\mu^2}
\Bigg\{
\frac{27}{y_4^2}
+\frac{1}{y_1^2}\nonumber\\
& +\frac{\kappa^2\bar v^2}{8}
\bigg[
9\frac{\left(2y_4-3y_1-3b_8\right)^2}{y_4^3}\nonumber\\
&-\frac{(y_1+b_8)^2}{y_1^3}
\bigg]
\Bigg\}.
\end{align}
Again, the first line reproduces the flat, vanishing-torsion part of the curved-space coefficient \eqref{eq:A2}, whereas the second line is a local off-shell contribution induced by $\bar v_\mu$.  The equality between the two unimodularity implementations is therefore an equality of the physical trace-log, not of the full gauge-fixed Hessians.

The structure of Eqs.~\eqref{eq:flat-final-axial-main}
and \eqref{eq:flat-final-vector-main} has a simple parametric
interpretation. With the exponential split, every metric-torsion
mixing vertex is proportional to $\kappa$. Hence, the leading
torsion-background-dependent correction is controlled by
$(\kappa\bar{\mathcal{A}})^2$ in the axial sector and by $(\kappa\bar v)^2$ in
the vector sector. These contributions are therefore suppressed
relative to the zero-background logarithmic divergence for weak
homogeneous torsion, $|\kappa\bar{\mathcal{A}}|\ll1$ or $|\kappa\bar v|\ll1$,
unless the coupling ratios multiplying them are parametrically large.

These flat-space checks give a useful diagnostic of the quantum relation between the Diff-invariant and unimodular theories beyond the strictly maximally symmetric, torsionless backgrounds used in the heat-kernel calculation.  The homogeneous background torsion generates metric-torsion mixing vertices which are absent for $\Tbar_\mu=0$, and therefore probes precisely the part of the quadratic action most sensitive to the absence of the trace mode in the unimodular theory.  The result shows that, once the unimodular gauge in the \diff-invariant theory is compared with the Landau \sdiff\ reduction of the genuinely unimodular theory, all torsion-dependent information is carried by the same physical determinant.  The remaining York, ghost, and group-volume factors are independent of the homogeneous background torsion and cannot modify the local $\Tbar^2$ logarithmic result.

At the same time, the argument remains a local off-shell check: it does not establish the equality of the two formulations for arbitrary torsionful curved backgrounds, where nonminimal operators and derivative-dependent background tensors may generate additional structures.  Within the present quadratic truncation, however, it provides nontrivial evidence that the unimodular constraint removes only the gauge/volume sector and leaves the physical torsion-loop contribution unchanged.

\section{Discussions and Conclusions}\label{sec:Conclusions}
In this work, we have investigated the relation between the Diff-invariant and unimodular formulations of the EFT-extended Poincar\'e gauge theory class in the presence of propagating torsion. Our analysis was carried out within a perturbative background-field framework and focused on the one-loop effective action for selected torsion sectors. The main purpose was to determine whether imposing the unimodular constraint modifies the local UV divergences generated by metric and torsion fluctuations.

For the class of backgrounds considered here, the logarithmically divergent part of the unimodular PGT effective action reproduces the corresponding local structure obtained in the Diff-invariant formulation. After decomposing the torsion into its irreducible components and accounting for the gauge and ghost sectors, the one-loop determinants yield the same curvature invariants with the same coefficients in the axial and vector torsion sectors. In this restricted but explicit sense, the unimodular constraint does not alter the one-loop renormalization of the local couplings resolved by our calculation.

This result should be understood with the appropriate qualifications. The equivalence established here concerns the local logarithmic divergences of the one-loop effective action on the backgrounds specified in the main text. It does not imply equality of the complete nonlocal effective actions or establish nonperturbative equivalence. Moreover, the curved-background calculation was performed on maximally symmetric backgrounds with vanishing torsion, for which the relevant fluctuation operators reduce to Laplace type.
General torsionful backgrounds may generate additional invariants and nonminimal metric-torsion operators that require generalized Schwinger-DeWitt or off-diagonal heat-kernel techniques~\cite{Barvinsky:1985an}.

The flat-space calculation of Sec.~\ref{sec:flat-space-consistency} provides a complementary consistency check in the presence of homogeneous torsion. There, Diff-invariant PGT in the unimodular gauge is compared with genuinely unimodular PGT reduced by its residual SDiff symmetry in Landau gauge. After gauge reduction, both formulations are governed by the same physical determinant, while the remaining York, ghost, and group-volume factors are
independent of the background torsion. The torsion-dependent logarithmic terms consequently agree at quadratic order. Since the homogeneous
background torsion is off-shell, this result remains a local consistency check and does not establish equivalence on arbitrary torsionful curved
backgrounds.

The role of the unimodular constraint is transparent in the present setting. It removes the metric trace fluctuation associated with the determinant of
the metric, while the torsion fluctuations remain unconstrained. The cosmological-volume contribution therefore has a different interpretation:
because the volume element is fixed, this term is field independent and does not affect the equations of motion. By contrast, the curvature-dependent divergences remain genuine local contributions and require the usual renormalization of the corresponding gravitational couplings. The agreement found here thus concerns the dynamical part of the one-loop divergences, not merely an additive constant in the effective action.

Our results extend previous observations on the perturbative relation between unimodular and standard metric gravity to a class of gauge theories of gravity with propagating torsion. This extension is nontrivial because PGT contains additional dynamical fields and couplings absent in purely metric gravity. In the axial and vector sectors considered here, these additional torsional degrees of freedom preserve the local one-loop structure of the Diff-invariant PGT result within the stated assumptions.

An important next step is to perform the computation on curved backgrounds with nonvanishing torsion. Such backgrounds generate additional invariants
built from curvature, torsion, and their covariant derivatives, together with nonminimal mixing between metric and torsion fluctuations.

Another natural extension is the densitized formulation of unimodular gravity \cite{deLeonArdon:2017qzg,Gielen:2018pvk}. There one writes
\begin{equation}
g_{\mu\nu}
=
\left(\frac{|\gamma|}{\omega^2}\right)^{-1/d}
\gamma_{\mu\nu},
\qquad
\det g_{\mu\nu}=\omega^2 ,
\end{equation}
so that the auxiliary metric $\gamma_{\mu\nu}$ carries an additional local Weyl redundancy, $\gamma_{\mu\nu}\mapsto e^{2\alpha(x)}\gamma_{\mu\nu}$. In a Riemann-Cartan setting, the relevant question is whether this Weyl sector factors out as a pure gauge mode once the torsion variables and the functional measure are treated consistently. If torsion is chosen to be Weyl invariant, one expects the densitized and constrained unimodular formulations to be equivalent after the Weyl gauge volume and Jacobians are included \cite{Sauro:2022chz}. If, instead, the torsion trace participates in the Weyl transformation, additional mixing and ghost determinants may arise. A separate analysis is required to settle this point.

A further direction is to enlarge the class of (extended) PGT actions and address the radiative stability of the propagating-torsion spectrum. The special coupling relations that remove ghost or tachyonic excitations at the classical level are not generically expected to define renormalization-group-invariant subspaces unless they are protected by additional symmetries or dynamical mechanisms \cite{Marzo:2021iok,Marzo:2024pyn}. A calculation resolving the complete dimension-four operator basis would therefore determine whether such classically healthy parameter subspaces are preserved by quantum corrections in either the Diff-invariant or unimodular formulation. If they are not, a quantum-mechanically consistent theory would require an additional mechanism for eliminating or consistently treating the unwanted modes, potentially along the lines of prescriptions investigated in higher-derivative gravity and related metric-affine constructions \cite{Anselmi:2020opi}. The present one-loop equivalence should consequently be understood independently of this separate question of quantum spectral consistency.

The inclusion of fermionic matter is especially relevant because Dirac fields couple directly to the torsion spin current and, under minimal coupling,
probe the axial torsion sector. Below the mass scale of a propagating torsion mode, torsion exchange generates effective four-fermion operators, see, e.g., \cite{Shaposhnikov:2020aen}. In the
quark sector, these operators can be combined with QCD-induced Fierz-rearranged four-quark interactions and evolved to the chiral-symmetry-breaking scale of QCD.  The related chiral crossover produces a vacuum-energy contribution, see, e.g., \cite{Alkofer:1988tc}. This provides a concrete setting in which to investigate a cosmological term induced by QCD and torsion. In unimodular PGT, the central question is whether this contribution only shifts the cosmological integration constant or also enters the running of curvature- and torsion-dependent couplings. Addressing it requires a consistent matching of the torsion and QCD scales, including nonperturbative information on the quark condensates.

Finally, the logarithmic divergences derived here provide the perturbative input for a renormalization-group analysis of unimodular Poincaré gauge theories. A natural next step is to compare the fixed-point structure and critical exponents of the Diff-invariant and unimodular formulations in enlarged classes of actions, adopting, for instance, nonperturbative functional renormalization-group techniques. Such an analysis may clarify whether an asymptotically safe unimodular Poincaré gauge theory of gravity can be realized.

\begin{acknowledgements}
We gratefully acknowledge Roberto Percacci for valuable comments and suggestions on the manuscript. A.F.V. is supported by a postdoctoral grant from FAPERJ in the Pós-doutorado Nota 10 program, under the grant No. 200.135/2025 and 200.136/2025. A.F.V. is also grateful to the Institute of Physics at the University of Graz for the hospitality extended to him and to the University of Graz for the financial support of his visit during the early stages of this work. A.D.P. acknowledges support from CNPq under the grant PQ-B (307623/2026-2) and from FAPERJ through the “Jovem Cientista do Nosso Estado” program (E-26/204.457/2025).
\end{acknowledgements}

\bigskip

{\bf Open Access Statement} --- For the purpose of open access, the authors have applied a Creative Commons Attribution (CC BY) licence  to any Author Accepted Manuscript version arising.

{\bf Research Data Access Statement} --- No data have been generated for this manuscript.
The software used to derive the Hessians will be made available on reasonable request.

\begin{widetext}
\appendix

\section{Trace of the heat-kernel coefficients}\label{app:trace_heatkernels}
In this section, we collect the traces of the local heat-kernel coefficients that enter the logarithmically divergent part of the one-loop effective action \cite{Avramidi:2000bm,Percacci:2017fkn,Buchbinder:2021wzv}. We restrict the evaluation to maximally symmetric curvature backgrounds and to vanishing background torsion. For each fluctuation sector, the relevant second-order operator is of Laplace type and can be written schematically as
\begin{equation}
	\Delta_s=\bar\Delta+\zeta_s \bar R+\mathbb{E}_s ,
\end{equation}
where $\bar\Delta=-\bar\nabla^2$ is the background Laplacian, $\zeta_s$ is a numerical coefficient depending on the spin and tensor structure of the fluctuation, and $\mathbb{E}_s$ denotes the non-derivative endomorphism part of the operator. Since we are interested in the logarithmic divergence in four dimensions, the relevant coefficient is $b_4[\Delta_s]$. We denote its trace over the corresponding spacetime indices by $\tr b_4[\Delta_s]$.
The general forms of $\tr b_4[\Delta_s]$ for each spin sector are
\begin{subequations}
	\begin{align}
		\tr b_4[\Delta_2]&=\frac{5}{2}\mathbb E_2^2
		+
		\frac{5}{6}\bar R\,\mathbb E_2
		\left(1+6\zeta_2\right)+\bar R^2
		\left(
		-\frac{1}{432}
		+
		\frac{5}{6}\zeta_2
		+
		\frac{5}{2}\zeta_2^2
		\right),\\
		\tr b_4[\Delta_1]&=
		\frac{3}{2}\mathbb E_1^2
		+
		\frac{1}{4}\bar R\,\mathbb E_1
		\left(12\zeta_1-1\right)+
		\bar R^2
		\left(
		-\frac{7}{1440}
		-
		\frac{1}{4}\zeta_1
		+
		\frac{3}{2}\zeta_1^2
		\right),\\
		\tr b_4[\Delta_0]&=
		\frac{1}{2}\mathbb E_0^2
		+
		\left(
		\zeta_0-\frac{1}{6}
		\right)
		\bar R\,\mathbb E_0+
		\bar R^2
		\left(
		\frac{29}{2160}
		-
		\frac{\zeta_0}{6}
		+
		\frac{\zeta_0^2}{2}
		\right).
	\end{align}
\end{subequations}
For the operators appearing in the axial and vector torsion sectors, we find	
\begin{subequations}
	\begin{align}
		\tr b_4[\Delta_{h^{\t\t}}]&=\frac{89}{432}\bar R^2,\\
		\tr b_4[\Delta_{a^\t}]&=\frac{27 m_1^4}{2(3 b_1+b_4)^2}
		-\frac{3(3 b_1-2 b_4)m_1^2 }{4(3 b_1+b_4)^2}\bar R-\frac{(63 b_1^2+312 b_1 b_4-38 b_4^2)}{1440(3 b_1+b_4)^2}\bar R^2\\
		\tr b_4[\Delta_{a^\l}]&=\frac{m_1^4}{2 b_1^2}
		-\frac{5 m_1^2 }{12 b_1}\bar R
		+\frac{373 }{4320} \bar R^2,\\
		\tr b_4[\Delta_{\varphi^\t}]&=\frac{27 y_2^2}{2 (3 b_8 + 3 y_1 + y_3)^2}+-\frac{3  y_2 (3 b_8 + 3 y_1 - 2 y_3)}{4 (3 b_8 + 3 y_1 + y_3)^2}\bar R\nonumber\\
		&\frac{1}{1440}\bigg(-7 + \frac{135 y_3^2}{(3 b_8 + 3 y_1 + y_3)^2} - \frac{90 y_3}{3 b_8 + 3 y_1 + y_3}\bigg)\bar R^2\\
		\tr b_4[\Delta_{\varphi^\l}]&=\frac{y_2^2}{2\,y_1^2} -\frac{\left(3b_8+5y_1\right)y_2}{12\,y_1^2} \bar R\frac{135 b_8^2+450 b_8 y_1+373 y_1^2}{4320\,y_1^2}\bar R^2\\
		\tr b_4[\Delta_{C^\t}]&=\frac{109}{720}\bar R^2,
	\end{align}
\end{subequations}
where the parameters \(y_i\) are defined in Eqs.~\eqref{eq:coefficients_vector1}, \eqref{eq:coefficients_vector2}, \eqref{eq:coefficients_vector3}, and \eqref{eq:coefficients_vector4}.

\section{Gauge-fixed flat-space Hessians}
\label{app:flat-four-hessians}

The quadratic actions displayed below are written as
\begin{equation}
	S^{(2)}=\frac12\int \dd^4x\,\mathcal H.
	\label{eq:app-half-factor-v5}
\end{equation}
We keep only the structures needed for the homogeneous-torsion check and use \(\Delta=-\partial^2\) in all flat-space Hessians. Since the metric background is flat and the background torsion is homogeneous, \(\Delta\) commutes with \(\partial_\mu\). Powers of \(\Delta\) have therefore been combined whenever they act on the same field.  Residual factors of \(\Delta^{-1/2}\) are kept only where they originate from the normalized longitudinal decomposition.  The standard Hessians contain the trace and the full \diff\ gauge sector, whereas the genuinely unimodular Hessians have no trace fluctuation and only the transverse \sdiff\ gauge sector.  This is the only distinction that survives before the Landau/unimodular reduction.

\subsection{Genuinely unimodular axial PGT}
\label{app:hessian-unimodular-axial-v6}

The genuinely unimodular axial Hessian contains no trace fluctuation.  It is convenient to decompose it according to powers of the homogeneous axial background,
\begin{equation}
	\mathcal H_{\uPGT,A}
	=\mathcal H_{\uPGT,A}^{(0)}+
	\mathcal H_{\uPGT,A}^{(\bar\A)}+
	\mathcal H_{\uPGT,A}^{(\bar\A^2)} .
	\label{eq:app-unimodular-axial-split-v6}
\end{equation}
The background-independent part is
\begin{align}
	\mathcal H_{\uPGT,A}^{(0)}
	={}&
	h^{{\rm TT}\,\alpha\beta}\Delta\hTT_{\alpha\beta}
	+\frac{2}{\alpha}\xi^\alpha\Delta^2\xi_\alpha
	-\frac38\sigma\Delta^3\sigma
	+\frac13m_1^2\aT_\alpha a^{{\rm T}\alpha}
	+\frac{3b_1+b_4}{9}a^{{\rm T}\alpha}\Delta\aT_\alpha
	\nonumber\\
	&+\frac{m_1^2}{3}\aL\aL
	+\frac{b_1}{3}\aL\Delta\aL .
	\label{eq:app-unimodular-axial-H0-v6}
\end{align}
The terms linear in \(\bar\A_\mu\) are
\begin{align}
	\mathcal H_{\uPGT,A}^{(\bar\A)}
	={}&
	-\frac{2\kappa m_1^2}{3}\hTT_{\alpha\beta}\bar\A^\alpha a^{{\rm T}\beta}
	-\frac{b_1\kappa}{6}\bar\A^\alpha a^{{\rm T}\beta}\Delta\hTT_{\alpha\beta}
	-\frac{b_1\kappa}{6}\hTT_{\alpha\beta}\bar\A^\alpha\Delta a^{{\rm T}\beta}
	\nonumber\\
	&-\frac{\kappa m_1^2}{3}\bar\A^\alpha\xi^\beta\partial_\alpha\aT_\beta
	-\frac{\kappa m_1^2}{3}\bar\A^\alpha a^{{\rm T}\beta}\partial_\alpha\xi_\beta
	+\frac{\kappa m_1^2}{3}\bar\A^\alpha\xi_\alpha\Delta^{1/2}\aL
	-\frac{\kappa m_1^2}{3}\bar\A^\alpha\aL\Delta^{1/2}\xi_\alpha
	\nonumber\\
	&-\frac{b_1\kappa}{3}\bar\A^\alpha\xi_\alpha\Delta^{3/2}\aL
	-\frac{b_1\kappa}{3}\bar\A^\alpha\aL\Delta^{3/2}\xi_\alpha
	-\frac{\kappa m_1^2}{12}\bar\A^\alpha\sigma\Delta\aT_\alpha
	-\frac{\kappa m_1^2}{12}\bar\A^\alpha\aT_\alpha\Delta\sigma
	\nonumber\\
	&-\frac{b_1\kappa}{24}\bar\A^\alpha\sigma\Delta^2\aT_\alpha
	-\frac{b_1\kappa}{24}\bar\A^\alpha\aT_\alpha\Delta^2\sigma
	+\frac{\kappa m_1^2}{4}\bar\A^\alpha\sigma\partial_\alpha\Delta^{1/2}\aL
	-\frac{\kappa m_1^2}{4}\bar\A^\alpha\aL\partial_\alpha\Delta^{1/2}\sigma
	\nonumber\\
	&+\frac{b_1\kappa}{8}\bar\A^\alpha\sigma\partial_\alpha\Delta^{3/2}\aL
	-\frac{b_1\kappa}{8}\bar\A^\alpha\aL\partial_\alpha\Delta^{3/2}\sigma .
	\label{eq:app-unimodular-axial-HA-v6}
\end{align}
The quadratic terms are
\begin{align}
	\mathcal H_{\uPGT,A}^{(\bar\A^2)}
	={}&
	\frac{\kappa^2m_1^2}{6}h^{{\rm TT}\,\alpha}{}_{\theta}\hTT_{\beta}{}^{\theta}\bar\A_\alpha\bar\A^\beta
	-\frac{b_1\kappa^2}{12}h^{{\rm TT}\,\theta\kappa}\bar\A^\alpha\bar\A^\beta\partial_\alpha\partial_\beta\hTT_{\theta\kappa}
	+\frac{b_1\kappa^2}{6}\hTT_\alpha{}^\theta\bar\A^\alpha\bar\A^\beta\Delta\hTT_{\beta\theta}
	\nonumber\\
	&+\frac{\kappa^2m_1^2}{6}\hTT_{\beta\theta}\bar\A^\alpha\bar\A^\beta\partial_\alpha\xi^\theta
	-\frac{\kappa^2m_1^2}{6}\bar\A^\alpha\bar\A^\beta\xi^\theta\partial_\beta\hTT_{\alpha\theta}
	+\frac{\kappa^2m_1^2}{24}\bar\A^\alpha\bar\A^\beta\sigma\Delta\hTT_{\alpha\beta}
	\nonumber\\
	&+\frac{\kappa^2m_1^2}{24}\hTT_{\alpha\beta}\bar\A^\alpha\bar\A^\beta\Delta\sigma
	+\frac{b_1\kappa^2}{24}\bar\A^\alpha\bar\A^\beta\sigma\Delta^2\hTT_{\alpha\beta}
	+\frac{b_1\kappa^2}{24}\hTT_{\alpha\beta}\bar\A^\alpha\bar\A^\beta\Delta^2\sigma
	\nonumber\\
	&-\frac{\kappa^2m_1^2}{6}\bar\A^\alpha\bar\A^\beta\xi^\theta\partial_\alpha\partial_\beta\xi_\theta
	+\frac{\kappa^2m_1^2}{6}\bar\A^\alpha\bar\A^\beta\xi_\alpha\Delta\xi_\beta
	+\frac{b_1\kappa^2}{3}\bar\A^\alpha\bar\A^\beta\xi_\alpha\Delta^2\xi_\beta
	\nonumber\\
	&-\frac{\kappa^2m_1^2}{12}\bar\A^\alpha\bar\A^\beta\sigma\partial_\beta\Delta\xi_\alpha
	+\frac{\kappa^2m_1^2}{12}\bar\A^\alpha\bar\A^\beta\xi_\alpha\partial_\beta\Delta\sigma
	-\frac{b_1\kappa^2}{8}\bar\A^\alpha\bar\A^\beta\sigma\partial_\beta\Delta^2\xi_\alpha
	\nonumber\\
	&+\frac{b_1\kappa^2}{8}\bar\A^\alpha\bar\A^\beta\xi_\alpha\partial_\beta\Delta^2\sigma
	-\frac{\kappa^2m_1^2}{12}\bar\A^\alpha\bar\A^\beta\sigma\partial_\alpha\partial_\beta\Delta\sigma
	-\frac{5b_1\kappa^2}{96}\bar\A^\alpha\bar\A^\beta\sigma\partial_\alpha\partial_\beta\Delta^2\sigma
	\nonumber\\
	&+\frac{\kappa^2m_1^2}{96}\bar\A^2\sigma\Delta^2\sigma
	+\frac{b_1\kappa^2}{96}\bar\A^2\sigma\Delta^3\sigma .
	\label{eq:app-unimodular-axial-HA2-v6}
\end{align}

\subsection{Diff-invariant axial PGT}
\label{app:hessian-standard-axial-v6}

The Diff-invariant axial Hessian is obtained from the unimodular axial one by restoring the trace fluctuation and the scalar part of the full \diff\ gauge sector.  With the conventions of Eq.~\eqref{eq:app-half-factor-v5},
\begin{equation}
	\mathcal H_{\PGT,A}
	=\mathcal H_{\uPGT,A}
	+\Delta\mathcal H_A^{(0)}
	+\Delta\mathcal H_A^{(\bar\A)}
	+\Delta\mathcal H_A^{(\bar\A^2)} .
	\label{eq:app-standard-axial-from-upgt-v6}
\end{equation}
Together with Eqs.~\eqref{eq:app-unimodular-axial-H0-v6}--\eqref{eq:app-unimodular-axial-HA2-v6}, the following three expressions give the full standard gauge-fixed Hessian.  The background-independent addition is
\begin{align}
	\Delta\mathcal H_A^{(0)}={}&
	\Lambda(\htr)^2
	+\frac{9}{8\alpha}\sigma\Delta^3\sigma
	+\frac38\left(-1+\frac{\beta}{\alpha}\right)
	\left(\sigma\Delta^2\htr+\htr\Delta^2\sigma\right)
	-\left(\frac38-\frac{\beta^2}{8\alpha}\right)\htr\Delta\htr .
	\label{eq:app-standard-axial-tr0-v6}
\end{align}
The terms linear in \(\bar\A_\mu\) that distinguish the Diff-invariant Hessian from the genuinely unimodular one are
\begin{align}
	\Delta\mathcal H_A^{(\bar\A)}={}&
	\frac{2\kappa m_1^2}{3}\bar\A^\alpha\xi^\beta\partial_\alpha\aT_\beta
	-\frac{2\kappa m_1^2}{3}\bar\A^\alpha\xi_\alpha\Delta^{1/2}\aL
	+\frac{\kappa m_1^2}{6}\htr\bar\A^\alpha\aT_\alpha
	-\frac{b_1\kappa}{24}\bar\A^\alpha\aT_\alpha\Delta\htr
	\nonumber\\
	&-\frac{b_1\kappa}{24}\htr\bar\A^\alpha\Delta\aT_\alpha
	-\frac{\kappa m_1^2}{12}\bar\A^\alpha\aL\partial_\alpha\Delta^{-1/2}\htr
	+\frac{\kappa m_1^2}{12}\htr\bar\A^\alpha\partial_\alpha\Delta^{-1/2}\aL
	\nonumber\\
	&+\frac{b_1\kappa}{24}\bar\A^\alpha\aL\partial_\alpha\Delta^{1/2}\htr
	-\frac{b_1\kappa}{24}\htr\bar\A^\alpha\partial_\alpha\Delta^{1/2}\aL .
	\label{eq:app-standard-axial-trA-v6}
\end{align}
The quadratic trace-sector addition is
\begin{align}
	\Delta\mathcal H_A^{(\bar\A^2)}={}&
	-\frac{\kappa^2m_1^2}{12}\htr\hTT_{\alpha\beta}\bar\A^\alpha\bar\A^\beta
	+\frac{b_1\kappa^2}{24}\hTT_{\alpha\beta}\bar\A^\alpha\bar\A^\beta\Delta\htr
	+\frac{b_1\kappa^2}{24}\htr\bar\A^\alpha\bar\A^\beta\Delta\hTT_{\alpha\beta}
	\nonumber\\
	&+\frac{\kappa^2m_1^2}{12}\bar\A^\alpha\bar\A^\beta\xi_\alpha\partial_\beta\htr
	-\frac{\kappa^2m_1^2}{12}\htr\bar\A^\alpha\bar\A^\beta\partial_\beta\xi_\alpha
	-\frac{b_1\kappa^2}{24}\bar\A^\alpha\bar\A^\beta\xi_\alpha\partial_\beta\Delta\htr
	\nonumber\\
	&+\frac{b_1\kappa^2}{24}\htr\bar\A^\alpha\bar\A^\beta\partial_\beta\Delta\xi_\alpha
	-\frac{\kappa^2m_1^2}{24}\bar\A^\alpha\bar\A^\beta\sigma\partial_\alpha\partial_\beta\htr
	-\frac{\kappa^2m_1^2}{24}\htr\bar\A^\alpha\bar\A^\beta\partial_\alpha\partial_\beta\sigma
	\nonumber\\
	&+\frac{b_1\kappa^2}{96}\bar\A^\alpha\bar\A^\beta\sigma\partial_\alpha\partial_\beta\Delta\htr
	+\frac{b_1\kappa^2}{96}\htr\bar\A^\alpha\bar\A^\beta\partial_\alpha\partial_\beta\Delta\sigma
	-\frac{\kappa^2m_1^2}{96}\bar\A^2\sigma\Delta\htr
	\nonumber\\
	&-\frac{\kappa^2m_1^2}{96}\htr\bar\A^2\Delta\sigma
	+\frac{b_1\kappa^2}{96}\bar\A^2\sigma\Delta^2\htr
	+\frac{b_1\kappa^2}{96}\htr\bar\A^2\Delta^2\sigma
	+\frac{\kappa^2m_1^2}{96}(\htr)^2\bar\A^2
	\nonumber\\
	&-\frac{b_1\kappa^2}{96}\htr\bar\A^\alpha\bar\A^\beta\partial_\alpha\partial_\beta\htr
	+\frac{b_1\kappa^2}{96}\htr\bar\A^2\Delta\htr .
	\label{eq:app-standard-axial-trA2-v6}
\end{align}

\subsection{Genuinely unimodular vector PGT}
\label{app:hessian-unimodular-vector-v5}

For the vector sector we use the abbreviations (cf.\ Eqs.\ \eqref{eq:coefficients_vector1}, \eqref{eq:coefficients_vector2} and \eqref{eq:coefficients_vector4})
\begin{equation}
	y_1=2b_1+3b_3,
	\quad y_2=2m_1^2+3m_3^2,
	\quad \Kp=y_1+b_8,
	\quad \Km=y_1-b_8,
	\quad y_4=3y_1+2b_4+b_7+3b_8 .
	\label{eq:app-vector-abbrev-v5}
\end{equation}
The unimodular vector Hessian is
\begin{equation}
	\mathcal H_{\uPGT,v}
	=\mathcal H_{\uPGT,v}^{(0)}+
	\mathcal H_{\uPGT,v}^{(\bar v)}+
	\mathcal H_{\uPGT,v}^{(\bar v^2)} .
	\label{eq:app-upgt-vector-split-v5}
\end{equation}
The background-independent part is
\begin{align}
	\mathcal H_{\uPGT,v}^{(0)}={}&
	h^{{\rm TT}\,\alpha\beta}\Delta h^{\rm TT}_{\alpha\beta}
	+\frac{2}{\alpha}\xi^\alpha\Delta^2\xi_\alpha
	-\frac38\sigma\Delta^3\sigma
	+\frac{2y_2}{3}\varphi^{{\rm T}\alpha}\varphi^\t_\alpha
	+\frac{2y_4}{9}\varphi^{{\rm T}\alpha}\Delta\varphi^\t_\alpha
	\nonumber\\
	&+\frac{2y_2}{3}\varphi^\l\varphi^\l
	+\frac{2y_1}{3}\varphi^\l\Delta\varphi^\l .
	\label{eq:app-upgt-vector-H0-v5}
\end{align}
The linear terms are
\begin{align}
	\mathcal H_{\uPGT,v}^{(\bar v)}={}&
	-\frac{2\kappa y_2}{3}h^{\rm TT}_{\alpha\beta}\bar v^\alpha\varphi^{{\rm T}\beta}
	-\frac{\kappa \Kp}{3}h^{\rm TT}_{\alpha\beta}\bar v^\alpha\Delta\varphi^{{\rm T}\beta}
	-\frac{\kappa \Kp}{3}\bar v^\alpha\varphi^{{\rm T}\beta}\Delta h^{\rm TT}_{\alpha\beta}
	\nonumber\\
	&+\frac{2\kappa y_2}{3}\bar v^\alpha\xi^\beta\partial_\alpha\varphi^\t_\beta
	-\frac{2\kappa y_2}{3}\bar v^\alpha\varphi^{{\rm T}\beta}\partial_\alpha\xi_\beta
	-\frac{2\kappa y_2}{3}\bar v^\alpha\xi_\alpha\Delta^{1/2}\varphi^\l
	-\frac{2\kappa y_2}{3}\bar v^\alpha\varphi^\l\Delta^{1/2}\xi_\alpha
	\nonumber\\
	&-\frac{2\kappa y_1}{3}\bar v^\alpha\xi_\alpha\Delta^{3/2}\varphi^\l
	-\frac{2\kappa y_1}{3}\bar v^\alpha\varphi^\l\Delta^{3/2}\xi_\alpha
	+\frac{\kappa y_2}{6}\bar v^\alpha\sigma\Delta\varphi^\t_\alpha
	-\frac{\kappa y_2}{6}\bar v^\alpha\varphi^\t_\alpha\Delta\sigma
	\nonumber\\
	&-\frac{\kappa \Kp}{12}\bar v^\alpha\sigma\Delta^2\varphi^\t_\alpha
	-\frac{\kappa \Kp}{12}\bar v^\alpha\varphi^\t_\alpha\Delta^2\sigma
	+\frac{\kappa y_2}{2}\bar v^\alpha\sigma\partial_\alpha\Delta^{1/2}\varphi^\l
	-\frac{\kappa y_2}{2}\bar v^\alpha\varphi^\l\partial_\alpha\Delta^{1/2}\sigma
	\nonumber\\
	&+\frac{\kappa \Km}{4}\bar v^\alpha\sigma\partial_\alpha\Delta^{3/2}\varphi^\l
	-\frac{\kappa \Km}{4}\bar v^\alpha\varphi^\l\partial_\alpha\Delta^{3/2}\sigma .
	\label{eq:app-upgt-vector-Hv-v5}
\end{align}
The quadratic terms are
\begin{align}
	\mathcal H_{\uPGT,v}^{(\bar v^2)}={}&
	\frac{\kappa^2y_2}{3}
	h^{{\rm TT}\,\alpha}{}_{\theta}h^{{\rm TT}}_{\beta}{}^{\theta}\bar v_\alpha\bar v^\beta
	-\frac{\kappa^2\Kp}{6}
	h^{{\rm TT}\,\theta\kappa}\bar v^\alpha\bar v^\beta
	\partial_\alpha\partial_\beta h^{\rm TT}_{\theta\kappa}
	+\frac{\kappa^2\Kp}{3}
	h^{{\rm TT}}_{\alpha}{}^{\theta}\bar v^\alpha\bar v^\beta\Delta h^{\rm TT}_{\beta\theta}
	\nonumber\\
	&-\frac{\kappa^2y_2}{3}
	h^{\rm TT}_{\beta\theta}\bar v^\alpha\bar v^\beta\partial_\alpha\xi^\theta
	-\frac{\kappa^2y_2}{3}
	\bar v^\alpha\bar v^\beta\xi^\theta\partial_\beta h^{\rm TT}_{\alpha\theta}
	+\frac{\kappa^2y_2}{12}
	h^{\rm TT}_{\alpha\beta}\bar v^\alpha\bar v^\beta\Delta\sigma
	\nonumber\\
	&+\frac{\kappa^2y_2}{12}
	\bar v^\alpha\bar v^\beta\sigma\Delta h^{\rm TT}_{\alpha\beta}
	+\frac{\kappa^2\Kp}{12}
	h^{\rm TT}_{\alpha\beta}\bar v^\alpha\bar v^\beta\Delta^2\sigma
	+\frac{\kappa^2\Kp}{12}
	\bar v^\alpha\bar v^\beta\sigma\Delta^2h^{\rm TT}_{\alpha\beta}
	\nonumber\\
	&-\frac{\kappa^2y_2}{3}\bar v^\alpha\bar v^\beta\xi^\theta
	\partial_\alpha\partial_\beta\xi_\theta
	+\frac{\kappa^2y_2}{3}\bar v^\alpha\bar v^\beta\xi_\alpha\Delta\xi_\beta
	+\frac{2\kappa^2y_1}{3}\bar v^\alpha\bar v^\beta\xi_\alpha\Delta^2\xi_\beta
	\nonumber\\
	&+\frac{\kappa^2y_2}{6}\bar v^\alpha\bar v^\beta\xi_\alpha\partial_\beta\Delta\sigma
	-\frac{\kappa^2\Km}{4}\bar v^\alpha\bar v^\beta\xi_\alpha\partial_\beta\Delta^2\sigma
	-\frac{\kappa^2y_2}{6}\bar v^\alpha\bar v^\beta\sigma\partial_\beta\Delta\xi_\alpha
	\nonumber\\
	&-\frac{\kappa^2\Km}{4}\bar v^\alpha\bar v^\beta\sigma\partial_\beta\Delta^2\xi_\alpha
	-\frac{\kappa^2y_2}{6}\bar v^\alpha\bar v^\beta\sigma\partial_\alpha\partial_\beta\Delta\sigma
	-\frac{\kappa^2(5y_1-13b_8)}{48}
	\bar v^\alpha\bar v^\beta\sigma\partial_\alpha\partial_\beta\Delta^2\sigma
	\nonumber\\
	&+\frac{\kappa^2y_2}{48}\bar v^2\sigma\Delta^2\sigma
	+\frac{\kappa^2\Kp}{48}\bar v^2\sigma\Delta^3\sigma .
	\label{eq:app-upgt-vector-Hv2-v5}
\end{align}

\subsection{Diff-invariant vector PGT}
\label{app:hessian-standard-vector-v5}

The Diff-invariant vector Hessian is most compactly written as the unimodular vector Hessian plus the trace and scalar-gauge sector,
\begin{equation}
	\mathcal H_{\PGT,v}
	=\mathcal H_{\uPGT,v}
	+\Delta\mathcal H_v^{(0)}
	+\Delta\mathcal H_v^{(\bar v)}
	+\Delta\mathcal H_v^{(\bar v^2)} .
	\label{eq:app-standard-vector-from-upgt-v5}
\end{equation}
This equation, together with Eqs.~\eqref{eq:app-upgt-vector-H0-v5}--\eqref{eq:app-upgt-vector-Hv2-v5} and the three expressions below, is the full gauge-fixed Hessian.  The background-independent trace/gauge contribution is
\begin{align}
	\Delta\mathcal H_v^{(0)}={}&
	\Lambda(\htr)^2
	+\frac{9}{8\alpha}\sigma\Delta^3\sigma
	+\frac38\left(-1+\frac{\beta}{\alpha}\right)
	\left(\sigma\Delta^2\htr+\htr\Delta^2\sigma\right)
	-\left(\frac38-\frac{\beta^2}{8\alpha}\right)\htr\Delta\htr .
	\label{eq:app-standard-vector-tr0-v5}
\end{align}
The terms linear in the vector background and involving the trace mode are
\begin{align}
	\Delta\mathcal H_v^{(\bar v)}={}&
	-\frac{\kappa y_2}{6}\htr\bar v^\alpha\varphi^\t_\alpha
	-\frac{\kappa \Kp}{12}\htr\bar v^\alpha\Delta\varphi^\t_\alpha
	+\frac{\kappa y_2}{6}\bar v^\alpha\varphi^\t_\alpha\htr
	-\frac{\kappa \Kp}{12}\bar v^\alpha\varphi^\t_\alpha\Delta\htr
	\nonumber\\
	&+\frac{\kappa y_2}{6}\htr\bar v^\alpha\partial_\alpha\Delta^{-1/2}\varphi^\l
	-\frac{\kappa (y_1+3b_8)}{12}\htr\bar v^\alpha\partial_\alpha\Delta^{1/2}\varphi^\l
	-\frac{\kappa y_2}{6}\bar v^\alpha\varphi^\l\partial_\alpha\Delta^{-1/2}\htr
	\nonumber\\
	&-\frac{\kappa (y_1+3b_8)}{12}\bar v^\alpha\varphi^\l\partial_\alpha\Delta^{1/2}\htr .
	\label{eq:app-standard-vector-trv-v5}
\end{align}
The quadratic trace-sector terms are
\begin{align}
	\Delta\mathcal H_v^{(\bar v^2)}={}&
	\frac{\kappa^2y_2}{12}h^{\rm TT}_{\alpha\beta}\bar v^\alpha\bar v^\beta\Delta\htr
	-\frac{\kappa^2\Kp}{12}h^{\rm TT}_{\alpha\beta}\bar v^\alpha\bar v^\beta\Delta^2\htr
	-\frac{\kappa^2y_2}{12}\htr\,h^{\rm TT}_{\alpha\beta}\bar v^\alpha\bar v^\beta
	\nonumber\\
	&+\frac{\kappa^2\Kp}{12}\htr\,\bar v^\alpha\bar v^\beta\Delta h^{\rm TT}_{\alpha\beta}
	+\frac{\kappa^2y_2}{6}\bar v^\alpha\bar v^\beta\xi_\alpha\partial_\beta\htr
	-\frac{\kappa^2(y_1+3b_8)}{12}\bar v^\alpha\bar v^\beta\xi_\alpha\partial_\beta\Delta\htr
	\nonumber\\
	&-\frac{\kappa^2y_2}{6}\htr\,\bar v^\alpha\bar v^\beta\partial_\alpha\xi_\beta
	+\frac{\kappa^2(y_1+3b_8)}{12}\htr\,\bar v^\alpha\bar v^\beta\partial_\alpha\Delta\xi_\beta
	-\frac{3}{32}\left(-4+\frac{\beta}{\alpha}\right)
	\left(\sigma\Delta^2\htr+\htr\Delta^2\sigma\right)
	\nonumber\\
	&-\frac{\kappa^2y_2}{12}\bar v^\alpha\bar v^\beta\sigma\partial_\alpha\partial_\beta\htr
	+\frac{\kappa^2(y_1+7b_8)}{48}\bar v^\alpha\bar v^\beta\sigma\partial_\alpha\partial_\beta\Delta\htr
	-\frac{\kappa^2y_2}{48}\bar v^2\sigma\Delta\htr
	\nonumber\\
	&+\frac{\kappa^2\Kp}{48}\bar v^2\sigma\Delta^2\htr
	-\frac{\kappa^2y_2}{12}\htr\,\bar v^\alpha\bar v^\beta\partial_\alpha\partial_\beta\sigma
	+\frac{\kappa^2(y_1+3b_8)}{48}\htr\,\bar v^\alpha\bar v^\beta\partial_\alpha\partial_\beta\Delta\sigma
	\nonumber\\
	&-\frac{\kappa^2y_2}{48}\htr\,\bar v^2\Delta\sigma
	+\frac{\kappa^2\Kp}{48}\htr\,\bar v^2\Delta^2\sigma
	+\frac{\kappa^2y_2}{48}(\htr)^2\bar v^2
	-\left(\frac38-\frac{\beta^2}{32\alpha}\right)\htr\Delta\htr
	\nonumber\\
	&+\frac{\kappa^2(-y_1+b_8)}{48}\htr\,\bar v^\alpha\bar v^\beta\partial_\alpha\partial_\beta\htr
	+\frac{\kappa^2\Kp}{12}\htr\,\bar v^2\Delta\htr .
	\label{eq:app-standard-vector-trv2-v5}
\end{align}

\section{Reduction to physical operators}
\label{app:flat-reduced-operators}

After the unimodular-gauge reduction in the Diff-invariant theory, or the $\textmd{SDiff}$ Landau-gauge reduction in the genuinely unimodular theory, the torsion-dependent determinant is evaluated on a reduced physical field space.  In the axial sector this multiplet is
\begin{equation}
	\Phi_{\rm phys}^{\rm axial}=
	\left(\hTT_{\mu\nu},s,\aT_\mu,\aL\right),
	\label{eq:app-axial-physical-multiplet-v11}
\end{equation}
while the vector sector is obtained by the replacement $(\aT_\mu,\aL)\mapsto(\varphi_\mu^{\rm T},\varphi^{\rm L})$.  The Landau-gauge decoupling of $\xi_\mu$ should not be interpreted as a removal of its functional integral.  In the genuinely unimodular theory, $\xi_\mu$ parametrizes the transverse gauge orbit,
\begin{equation}
	\delta_{\epsilon^{\rm T}}\xi_\mu=\epsilon_\mu^{\rm T},
	\qquad \partial^\mu\epsilon_\mu^{\rm T}=0,
	\label{eq:app-xi-sdiff-transformation-v11}
\end{equation}
and may be fixed by the transverse condition
\begin{equation}
	F_\mu^{\rm T}[h;\bar g]
	=\sqrt{2}\left(\delta_\mu{}^\nu-\partial_\mu\Delta^{-1}\partial^\nu\right)
	\partial^\rho h_{\nu\rho}
	=-\sqrt{2}\Delta\xi_\mu .
	\label{eq:app-upgt-transverse-gauge-condition-v11}
\end{equation}
This gives
\begin{equation}
	S_{\rm gf}^{\uPGT}=\frac{1}{\alpha}\int\dd^4x\,\xi_\mu\Delta^2\xi^\mu,
	\qquad
	S_{\rm gh}^{\uPGT}=-\sqrt{2}\int\dd^4x\,\bar C_{\rm T}^{\mu}\Delta C_\mu^{\rm T},
	\label{eq:app-upgt-gf-ghost-v11}
\end{equation}
so that $Z_{\rm gh}^{\uPGT}=\Det_{1{\rm T}}(\Delta)$.  For a homogeneous background torsion $\Tbar_\mu$, the quadratic form has the schematic block structure
\begin{equation}
	S_\alpha^{(2)}=\frac12
	\begin{pmatrix}\Phi_{\rm phys}&\xi\end{pmatrix}
	\begin{pmatrix}
		\mathcal O_{\rm phys}(\Tbar)&\mathcal V_{\Phi\xi}(\Tbar)\\
		\mathcal V_{\xi\Phi}(\Tbar)&\alpha^{-1}\Delta^2
	\end{pmatrix}
	\begin{pmatrix}\Phi_{\rm phys}\\ \xi\end{pmatrix} .
	\label{eq:app-full-block-hessian-v11}
\end{equation}
Using the block-determinant decomposition of Eq.~\eqref{eq:app-full-block-hessian-v11}, the mixing contribution involving $\xi_\mu$ vanishes in the Landau limit,
\begin{equation}
	\mathcal V_{\Phi\xi}(\alpha^{-1}\Delta^2)^{-1}\mathcal V_{\xi\Phi}
	=\alpha\,\mathcal V_{\Phi\xi}\Delta^{-2}\mathcal V_{\xi\Phi}
	\xrightarrow{\alpha\to0}0,
	\label{eq:app-xi-schur-landau-v11}
\end{equation}
so that the physical determinant is unaffected. The Gaussian integration over $\xi_\mu$ nevertheless yields
\begin{equation}
	Z_\xi=
	\Det_{1{\rm T}}(\alpha^{-1}\Delta^2)^{-1/2}
	=\mathcal N_\alpha\Det_{1{\rm T}}^{-1}(\Delta),
	\label{eq:app-xi-landau-determinant-v11}
\end{equation}
with $\mathcal N_\alpha$ background independent.  This cancels the transverse ghost determinant.  The remaining flat traceless-York factor is
\begin{equation}
	J_Y^{\uPGT}=\Det_{1{\rm T}}^{1/2}(\Delta)\Det_0(\Delta),
	\qquad
	(d\sigma)=\Det_0^{-1}(\Delta)(ds),
	\label{eq:app-flat-york-factor-v11}
\end{equation}
leaving $J_{\rm red}=\Det_{1{\rm T}}^{1/2}(\Delta)$.  After the metric-independent $\textmd{SDiff}$ group-volume normalization is accounted for, all residual determinants from gauge coordinates, ghosts, York factors, and group volume are independent of $\Tbar_\mu$.  They therefore do not contribute to the $\bar\A^2$ or $\bar v^2$ logarithmic contributions. 

\subsection{Axial sector}
\label{app:reduced-axial-v5}

In the basis $\Phi_{\rm phys}^{\rm axial}=(\hTT_{\mu\nu},s,\aT_\mu,\aL)$, define
\begin{equation}
	B=3b_1+b_4,
	\qquad f(p)=b_1\,p^2+2m_1^2.
\end{equation}
Suppressing tensor indices for notational simplicity, the free operator and its inverse are given, respectively, by
\begin{align}
	\mathcal O_0^{\rm axial}(p)
	&=\operatorname{diag}\left(
	p^2\PiTT,-\frac38 p^2,\frac{B\,p^2+3m_1^2}{9}\PT,
	\frac{b_1\,p^2+m_1^2}{3}
	\right),
	\label{eq:axial-O0-v5}\\
	\mathcal G_0^{\rm axial}(p)
	&=\operatorname{diag}\left(
	\frac{\PiTT}{p^2},-\frac{8}{3p^2},
	\frac{9\PT}{B\,p^2+3m_1^2},
	\frac{3}{b_1\,p^2+m_1^2}
	\right),
	\label{eq:axial-G0-v5}
\end{align}
where we define the transverse vector projector and the
transverse-traceless tensor projector in momentum space as
\begin{align}
	P^{\rm T}_{\mu\nu}(p)
	&=
	\delta_{\mu\nu}
	-\frac{p_\mu p_\nu}{p^2},
	\label{eq:app-transverse-projector}
	\\
	\Pi^{\rm TT}_{\mu\nu,\rho\sigma}(p)
	&=
	\frac{1}{2}
	\left(
	P^{\rm T}_{\mu\rho}P^{\rm T}_{\nu\sigma}
	+
	P^{\rm T}_{\mu\sigma}P^{\rm T}_{\nu\rho}
	\right)
	-\frac{1}{3}
	P^{\rm T}_{\mu\nu}P^{\rm T}_{\rho\sigma}.
	\label{eq:app-tt-projector}
\end{align}
In four dimensions, these projectors satisfy
\begin{equation}
	\tr P^{\rm T}=3,
	\qquad
	\tr \Pi^{\rm TT}=5,
	\qquad
	\Pi^{\rm TT}_{\mu\lambda,\nu}{}^{\lambda}
	=\frac{5}{3}P^{\rm T}_{\mu\nu}.
	\label{eq:app-projector-identities}
\end{equation}
The nonzero $\Order(\bar\A)$ blocks are
\begin{align}
	(\mathcal N_{ha})^{\mu\nu,\rho}
	&=-\frac{\kappa f(p)}{6}\bar\A^\mu\delta^{\nu\rho},
	&
	(\mathcal N_{sa})^\rho
	&=-\frac{\kappa f(p)}{24}\bar\A^\rho,
	\label{eq:axial-N1-trans-v5}\\
	\mathcal N_{sL}
	&=-\frac{i\kappa f(p)}{8\sqrt{p^2}}(\bar\A\cdot p),
	&
	\mathcal N_{Ls}&=-\mathcal N_{sL},
	\label{eq:axial-N1-long-v5}
\end{align}
with the Hermitian-conjugate reverse blocks understood.  The diagonal $\Order(\bar\A^2)$ insertions entering the single trace are
\begin{align}
	h^{\rm TT}Q^{\rm A}_{hh}h^{\rm TT}
	={}&\kappa^2\left[
	\frac16(b_1\,p^2+m_1^2)h^{\rm TT}_{\mu\lambda}h^{{\rm TT}\lambda}{}_{\nu}\bar\A^\mu\bar\A^\nu
	+\frac{b_1}{12}(\bar\A\cdot p)^2h^{\rm TT}_{\mu\nu}h^{{\rm TT}\mu\nu}
	\right],
	\label{eq:axial-Qhh-v5}\\
	sQ^{\rm A}_{ss}s
	={}&\kappa^2s^2\left[
	\frac{m_1^2}{12p^2}(\bar\A\cdot p)^2+
	\frac{5b_1}{96}(\bar\A\cdot p)^2+
	\frac{m_1^2}{96}\bar\A^2+
	\frac{b_1\,p^2}{96}\bar\A^2
	\right].
	\label{eq:axial-Qss-v5}
\end{align}
The off-diagonal $h^{\rm TT}$-$s$ entry is part of $\mathcal N_2$ but has vanishing block trace in $\Tr(\mathcal G_0\mathcal N_2)$.

\subsection{Vector sector}
\label{app:reduced-vector-v5}

In the basis $\Phi_{\rm phys}^{\rm vec}=(\hTT_{\mu\nu},s,\varphi^\t_\mu,\varphi^\l)$,
\begin{align}
	\mathcal O_0^{\rm vec}(p)
	&=\operatorname{diag}\left(
	p^2\PiTT,-\frac38 p^2,
	\frac{2}{9}(y_4\,p^2+3y_2)\PT,
	\frac{2}{3}(y_1\,p^2+y_2)
	\right),
	\label{eq:vector-O0-v5}\\
	\mathcal G_0^{\rm vec}(p)
	&=\operatorname{diag}\left(
	\frac{\PiTT}{p^2},-\frac{8}{3p^2},
	\frac{9\PT}{2(y_4\,p^2+3y_2)},
	\frac{3}{2(y_1\,p^2+y_2)}
	\right).
	\label{eq:vector-G0-v5}
\end{align}
The nonzero $\Order(\bar v)$ blocks are
\begin{align}
	(\mathcal N_{h\varphi})^{\mu\nu,\rho}
	&=-\frac{\kappa}{3}(\Kp\,p^2+2y_2)\bar v^\mu\delta^{\nu\rho},
	&
	(\mathcal N_{s\varphi})^\rho
	&=-\frac{\kappa}{12}(\Kp\,p^2+2y_2)\bar v^\rho,
	\label{eq:vector-N1-trans-v5}\\
	\mathcal N_{sL}
	&=+\frac{i\kappa}{4\sqrt{p^2}}(\Km\,p^2+2y_2)(\bar v\cdot p),
	&
	\mathcal N_{Ls}&=-\mathcal N_{sL}.
	\label{eq:vector-N1-long-v5}
\end{align}
The diagonal $\Order(\bar v^2)$ insertions are
\begin{align}
	h^{\rm TT}Q^{\rm v}_{hh}h^{\rm TT}
	={}&\kappa^2\left[
	\frac13(\Kp\,p^2+y_2)h^{\rm TT}_{\mu\lambda}h^{{\rm TT}\lambda}{}_{\nu}\bar v^\mu\bar v^\nu
	+\frac{\Kp}{6}(\bar v\cdot p)^2h^{\rm TT}_{\mu\nu}h^{{\rm TT}\mu\nu}
	\right],
	\label{eq:vector-Qhh-v5}\\
	sQ^{\rm v}_{ss}s
	={}&\kappa^2s^2\left[
	\frac{y_2}{6p^2}(\bar v\cdot p)^2
	+\frac{5y_1-13b_8}{48}(\bar v\cdot p)^2
	+\frac{y_2}{48}\bar v^2+\frac{\Kp\,p^2}{48}\bar v^2
	\right].
	\label{eq:vector-Qss-v5}
\end{align}
The $h^{\rm TT}$-$s$ entry again drops from the single trace.

\section{Momentum integrals and logarithmic contributions}
\label{app:flat-momentum-integrals}

The tensor reductions in momentum integrals used in both sectors are\footnote{Here, a remark about the validity of these tensor reductions is in order. First, as discussed in the main text in the present setting we assume $\Tbar^2\not=0$ without breaking isotropy or homogeneity. Furthermore, 
	although a nonzero constant background $\Tbar_\mu$ breaks the $O(4)$ invariance of the chosen background configuration by selecting a preferred direction, the tensor reductions used in the loop integrals are nevertheless still valid for the used purpose. They rely only on the $O(4)$-invariance of the flat momentum measure and of the zeroth-order propagator, treating $\Tbar_\mu$ as an external \textit{spurion}. 
}
\begin{align}
	\intp p_\mu p_\nu f(p)&=\frac{\delta_{\mu\nu}}{4}\intp p^2 f(p),
	&
	\intp(\Tbar\cdot p)^2f(p)&=\frac{\Tbar^2}{4}\intp p^2 f(p),
	\label{eq:tensor-reductions-v5}\\
	\intp \Tbar^\mu \PT_{\mu\nu}\Tbar^\nu f(p)
	&=\frac{3\Tbar^2}{4}\intp f(p),
	&
	\intp\frac{1}{(p^2)^2}&=\frac{1}{16\pi^2}\log\frac{\Lambda_{\rm UV}^2}{\mu^2}.
	\label{eq:basic-log-v5}
\end{align}

\subsection{Axial sector}

The background-independent logarithmic part is
\begin{equation}
	\Gamma_{\log}^{(0),\rm axial}
	=-\frac{V_4m_1^4}{64\pi^2}
	\left(\frac{27}{B^2}+\frac{1}{b_1^2}\right)
	\log\frac{\Lambda_{\rm UV}^2}{\mu^2},
	\qquad B=3b_1+b_4 .
	\label{eq:axial-zero-residue-v5}
\end{equation}
The direct insertion is
\begin{equation}
	\Gamma_{\rm dir}^{(\bar\A^2)}
	=\frac12\Tr(\mathcal G_0^{\rm axial}\mathcal N_2^{\rm axial})
	=\kappa^2\bar\A^2V_4\intp
	\left(\frac{b_1}{8}+\frac{m_1^2}{16p^2}\right),
	\label{eq:axial-direct-v5}
\end{equation}
which has no logarithmic contribution.  The exchange channels combine to
\begin{equation}
	\Gamma_{\rm exch}^{(\bar\A^2)}
	=\frac{\kappa^2\bar\A^2V_4}{64}\intp
	\left[
	-9\frac{f(p)^2}{p^2(B\,p^2+3m_1^2)}+
	\frac{f(p)^2}{p^2(b_1\,p^2+m_1^2)}
	\right].
	\label{eq:axial-exchange-v5}
\end{equation}
The UV asymptotics are obtained from the large-momentum expansions
\begin{align}
	\frac{f(p)^2}{p^2(B\,p^2+3m_1^2)}
	&=\cdots+
	\frac{m_1^4(2B-3b_1)^2}{B^3}\frac{1}{(p^2)^2}+
	\Order(p^{-6}),
	\label{eq:axial-large-trans-v5}\\
	\frac{f(p)^2}{p^2(b_1\,p^2+m_1^2)}
	&=\cdots+
	\frac{m_1^4}{b_1}\frac{1}{(p^2)^2}+
	\Order(p^{-6}).
	\label{eq:axial-large-long-v5}
\end{align}
This gives Eq.~\eqref{eq:flat-final-axial-main}.

\subsection{Vector sector}

The background-independent logarithmic part is
\begin{equation}
	\Gamma_{\log}^{(0),\rm vec}
	=-\frac{V_4y_2^2}{64\pi^2}
	\left(\frac{27}{y_4^2}+\frac{1}{y_1^2}\right)
	\log\frac{\Lambda_{\rm UV}^2}{\mu^2}.
	\label{eq:vector-zero-residue-v5}
\end{equation}
The direct insertion is
\begin{equation}
	\Gamma_{\rm dir}^{(\bar v^2)}
	=\kappa^2\bar v^2V_4\intp
	\left(\frac{2y_1+3b_8}{8}+\frac{y_2}{8p^2}\right),
	\qquad
	\Gamma_{\rm dir,log}^{(\bar v^2)}=0.
	\label{eq:vector-direct-v5}
\end{equation}
The exchange contributions are
\begin{align}
	\Gamma^{(\bar v^2)}_{h\varphi}
	&=-\kappa^2\bar v^2V_4\intp
	\frac{5}{16}\frac{(\Kp\,p^2+2y_2)^2}{p^2(y_4\,p^2+3y_2)},
	\label{eq:vector-exchange-hv-v5}\\
	\Gamma^{(\bar v^2)}_{s\varphi}
	&=+\kappa^2\bar v^2V_4\intp
	\frac{1}{32}\frac{(\Kp\,p^2+2y_2)^2}{p^2(y_4\,p^2+3y_2)},
	\label{eq:vector-exchange-sv-v5}\\
	\Gamma^{(\bar v^2)}_{sL}
	&=+\kappa^2\bar v^2V_4\intp
	\frac{1}{32}\frac{(\Km\,p^2+2y_2)^2}{p^2(y_1\,p^2+y_2)}.
	\label{eq:vector-exchange-sl-v5}
\end{align}
Their logarithmic contribution is
\begin{equation}
	\Gamma_{\log}^{(\bar v^2),\rm vec}
	=\frac{\kappa^2\bar v^2V_4y_2^2}{2048\pi^2}
	\left[
	-36\frac{(2y_4-3\Kp)^2}{y_4^3}+
	4\frac{\Kp^2}{y_1^3}
	\right]
	\log\frac{\Lambda_{\rm UV}^2}{\mu^2}.
	\label{eq:vector-v2-residue-v5}
\end{equation}
Together with Eq.~\eqref{eq:vector-zero-residue-v5}, this yields Eq.~\eqref{eq:flat-final-vector-main}.

\end{widetext}

\begingroup
\allowdisplaybreaks

\bibliographystyle{apsrev4-1}

\bibliography{refs}

\end{document}